\documentclass[journal,draft,onecolumn,12pt]{IEEEtran}
\usepackage{amsfonts,color,morefloats}
\usepackage{amssymb,amsmath,latexsym,amsthm}
\usepackage{tikz,tabu}
\usepackage[thicklines]{cancel}

\newcommand{\Z}{\mathbb{{Z}}}

\newcommand{\bv}{{\mathbf{v}}}

\newtheorem{theorem}{Theorem}
\newtheorem{lemma}[theorem]{Lemma}

\newtheorem{corollary}[theorem]{Corollary}

\newtheorem{problem}{Open Problem}

\begin{document}

\title{A Survey on Codes from Simplicial Complexes \thanks{
 This work was sponsored by the National Natural Science Foundation of China under Grant Number 62372247 and 12101326, the
Natural Science Foundation of Jiangsu Province under Grant Number BK20210575, and the China Postdoctoral Science Foundation under Grant Number 2023M740958.}

}

\author{Yansheng Wu,\thanks{Y. Wu is with the School of Computer Science, Nanjing University of Posts and Telecommunications, Nanjing
210023, P. R. China. Email: 
yanshengwu@njupt.edu.cn} 
\and Chao Li,\thanks{C. Li is with the School of Computer Science, Nanjing University of Posts and Telecommunications, Nanjing, 210023, P. R. China. Email: 1022041222@njupt.edu.cn }
\and Jong Yoon Hyun\thanks{J. Y. Hyun  is with the Konkuk University, Glocal Campus, 268 Chungwon-daero Chungju-si Chungcheongbuk-do 27478, South Korea. Email: hyun33@kku.ac.kr}

}

\date{\today}
\maketitle



\begin{abstract}


In the field of mathematics, a purely combinatorial equivalent to a simplicial complex, or more generally, a down-set, is an abstract structure known as a family of sets. This family is closed under the operation of taking subsets, meaning that every subset of a set within the family is also included in the family. The purpose of this paper is two-fold. Firstly, it aims to present a comprehensive survey of recent results in the field. This survey intends to provide an overview of the advancements made in codes constructed from simplicial complexes. Secondly, the paper seeks to propose open problems that are anticipated to stimulate further research in this area. By highlighting these open problems, the paper aims to encourage and inspire future investigations and developments in the field of codes derived from simplicial complexes.
\end{abstract}






\section{Introduction}

In mathematics, a simplicial complex is a set composed of points, line segments, triangles, and their $n$-dimensional counterparts. The purely combinatorial counterpart to a simplicial complex is  abstract, which is a family of sets closed under taking subsets; i.e., every subset of a set in the family is also in the family.

Adamaszek \cite{A} proved formulas for two-variable generating functions associated with simplicial complexes by applying the inclusion–exclusion principle.  Chang and Hyun \cite{CH} introduced simplex complexes as a mathematical method in the construction of error-correcting codes, and they also found the first infinite family of minimal linear codes violating Ashikhmin-Barg's condition. Since then, many scholars have applied this method in various finite rings and finite fields over the past few years, resulting in numerous excellent codes. Among them are many codes that satisfy the Griesmer Bound and the Sphere Packing Bound. Hence, in this paper, we will provide a survey on codes from simplicial complexes and present some open problems.

\section{Preliminaries}

In this section, we will introduce some concepts and results, which will be needed.

\subsection{Two Constructions of Linear Codes}

Let $m$ be a positive integer, $q$ be a prime power, and $(V_m,\cdot)$ be an $m$-dimensional vector space over $\Bbb F_q$, where $\cdot$ denotes an inner product on $V_m$. For a linear code of length $n$ over $\Bbb F_q$, there is a generic construction as follows:
\begin{equation}\label{eq2.1}
\mathcal{C}_D = \{(x\cdot d_1, x\cdot d_2, \ldots, x\cdot d_n): x\in V_m\},
\end{equation}
where $D=\{d_1, \ldots, d_n\} \subseteq V_m$. The set $D$ is called the defining set of the code $\mathcal{C}_{D}$. By making appropriate choices for the set $D$, the code $\mathcal{C}_{D}$ can exhibit favorable parameters.

$(1)$ When $V_m=\Bbb F_{q^m}$, $x\cdot y=\text{Tr}_{q^{m}/q}(xy)$ for $x,y\in \Bbb F_{q^m}$, and $\text{Tr}_{q^{m}/q}$ is the trace function from $\Bbb F_{q^m}$ to $\Bbb F_{q}$. In this case, the corresponding code $\mathcal{C}_D$ in Equation \eqref{eq2.1} is called a trace code over $\Bbb F_q$. This generic construction was first introduced by Ding \cite{D1}. We will not deal with this construction in this paper.

$(2)$ When $V_m=\Bbb F_q^m$, ${\bf x}\cdot {\bf y}=\sum_{i=1}^mx_iy_i$ for ${\bf x}=(x_1, \ldots, x_m)$, ${\bf y}=(y_1, \ldots, y_m)\in \Bbb F_q^m$. This standard construction in Equation \eqref{eq2.1} can also be found in \cite{HP}. The construction in Equation \eqref{eq2.1} can also be generalized to finite rings and has been widely employed in various papers, including \cite{WZY, WLX, LS, SS, SL2, LS2, ZW, SS5, SLP}.

Another construction of binary linear codes is based on Boolean functions. Let $f$ be a non-linear $n$-variable Boolean function with $f({\bf 0}) = 0$. A binary linear code $\mathcal{C}_f$ associated with $f$ is defined  as follows:

\begin{equation}\label{eq2.2}
\mathcal{C}_f = \{ c_f(s,{\bf u}) = (sf(x) + {\bf u}\cdot x )_{x \in \Bbb Z_2^{n*}} : s\in \Bbb Z_2, {\bf u} \in \Bbb Z_2^n \},
\end{equation}
where $X^*$ denotes the set of nonzero elements of a set $X$, and $\Z_2$ is a ring of integers modular $2$.

\subsection{Three Different Metrics for Codes}

In this subsection, we will list some different metrics for codes.

\begin{itemize}

\item {\bf Hamming weight:}
Let $n$ be a positive integer and $\Bbb F_q^n$ denote the vector space of all $n$-tuples over the finite field $\Bbb F_q$. The Hamming distance $d({\bf x, y})$ between two vectors ${\bf x, y }\in \Bbb F_q^n$ is defined to be the number of coordinates in which ${\bf x}$ and ${\bf y}$ differ, and the Hamming weight of a vector is the number of symbols that are different from the zero vector. We call $\mathcal{C}$ an $[n,k,d]$ linear code over $\Bbb F_q$ if $\mathcal{C}$ is a subspace of $\Bbb F_q^n$ of dimension $k$ with minimum Hamming distance $d$. We also call the vectors in $\mathcal{C}$ codewords. Let $A_i$ denote the number of codewords in $\mathcal C$ with Hamming weight $i$. The weight enumerator of $\mathcal C$ is defined by $1+A_1z+A_2z^2+\cdots+A_nz^n$. The sequence $(1, A_1, A_2, \ldots, A_n)$ is called the weight distribution of $\mathcal C$. A code $\mathcal{C}$ is $t$-weight if the number of nonzero $A_{i}$ in the sequence $(A_1, A_2, \ldots, A_n)$ is equal to $t$.

\item {\bf Lee weight and Gray map:}
Let $\mathbb{Z}_4$ be a ring of integers modulo 4, $n$ be a positive integer, and $\mathbb{Z}_4^n$ be the set of $n$-tuples over $\mathbb{Z}_4$. The Lee weight of a vector ${\bf u}$ of length $n$ over $\mathbb{Z}_4$ is defined to be the Hamming weight of its Gray image as follows:
$$\mbox{wt}_{L}(\mathbf{u}) = \mbox{wt}_{L}(\mathbf{a}+2\mathbf{b}) = \mbox{wt}_{H}(\mathbf{b}) + \mbox{wt}_{H}(\mathbf{a}+\mathbf{b}).$$
The Lee distance of $\mathbf{x},\mathbf{y}\in \mathbb{Z}_4^n$ is defined as $\mbox{wt}_L(\mathbf{x-y})$. From \cite[Theorem 3.1]{WQ}, the Gray map is an isometry from $(\mathbb{Z}_4^n, d_L)$ to $(\mathbb{Z}_2^{2n}, d_H)$ and is a weight-preserving map.
Assume that a quaternary code $\mathcal C$ has Lee weights ${i_1, \ldots, i_k}$ and there are $A_i$ codewords in a quaternary $\mathcal C$ with the Lee weight $i$. Then $(1, A_{i_1}, \ldots, A_{i_k})$ and $1+A_{i_1}z^{i_1}+\cdots +A_{i_k}z^{i_k}\in \mathbb Z[z]$ are called Lee weight distribution and Lee weight enumerator of $\mathcal C$, respectively.

There are different Gray maps from $\Bbb Z_{2^k} \to \Bbb Z_{2}^{2^{k-1}}$. Codes constructed over rings of size four, such as $\Bbb Z_4$ (\cite{ZWY,WLZ}), $\Bbb F_4$ (\cite{WLX, ZW}), and $\Bbb F_2 + u\Bbb F_2$ (\cite{WZY, SLP}), have been extensively studied. The Gray map has been utilized to construct numerous codes over different rings, including $\Bbb F_2[u] \backslash \langle u^k \rangle$ \cite{LS2}, $\Bbb Z_2 [u] \backslash \langle u^3 - u \rangle$ \cite{SS2}, $\mathbb{Z}_2[u]$ \cite{KM}, $\Bbb Z_p\Bbb Z_p[u]$ \cite{WS}, and non-commutative and non-unital ring $E$ \cite{SS4} and commutative and non-unital ring $I$ \cite{SS3}. These rings provide a versatile framework for code construction, and when transformed under the Gray image, the resulting codes may exhibit good parameters.

\item {\bf Homogeneous weight:}
The homogeneous weight of a vector $\bf a$ of length $m$ over a ring $R$ is defined to be the Hamming weight of its Gray image as $\mbox{wt}_{hom}({\bf a}) = \mbox{wt}_{hom}(a_0 + a_1u + \cdots + a_{k-1}u^{k-1}) = \sum_{i=0}^{2^{k-1}}b_i $, where $b_{2i+\varepsilon} = a_{k-1} + \sum_{l=1}^{k-2}p_{l-1}(i)a_{l} + \varepsilon a_0$ for $0 \leq i \leq 2^{k-2} - 1$, $0 \leq \varepsilon \leq 1$. The homogeneous distance $d_{hom}({\bf a, b})$ of ${\bf a, b} \in R^m$ is defined as $\mbox{wt}_{hom}({\bf a}- {\bf b})$. Thus the Gray map is an isometry from $(R^m, d_{hom})$ to $(\Bbb F_2^{2^{k-1}m}, d_{H})$. Note that when $k = 2$, the homogeneous weight is just the Lee weight \cite{WZY}.

\end{itemize}

\subsection{Minimal Linear Codes}

For two vectors ${\bf u,v}\in \Bbb F_2^n$, we say that ${\bf u}$ covers ${\bf v}$ if $\mathrm{supp}({\bf v})\subseteq\mathrm{supp}({\bf u})$. A nonzero codeword ${\bf u}$ in a linear code $\mathcal{C}$ is said to be {\bf minimal} if ${\bf u}$ covers the zero vector and ${\bf u}$ itself but no other codewords in the code $\mathcal{C}$. A linear code $\mathcal{C}$ is said to be {\bf minimal} if every nonzero codeword in $\mathcal{C}$ is minimal.

A sufficient condition for a linear code to be minimal is due to Ashikhmin-Barg. We present the sufficient condition in detail.

\begin{lemma}\label{lem2.1} {\rm (Ashikhmin-Barg \cite{AB})
Let $\mathcal{C}$ be a linear code over $\mathbb{F}_q$ with $\mbox{wt}_{min}$ and $\mbox{wt}_{max}$ as minimum and maximum weights,  respectively, of its non-zero codewords. If
$$\frac{\mbox{wt}_{min}}{\mbox{wt}_{max}} > \frac{q-1}{q},$$ 
then $\mathcal{C}$ is minimal.
}
\end{lemma}

\subsection{Subfield Codes}

Suppose that $\mathcal{C}$ is a $k$-dimensional code of length $n$ over $\Bbb F_{q^l}$, generated by $G$, and $\mathcal{B}$ is an ordered basis of $\Bbb F_{q^l}$ over $\Bbb F_q$. The code $\mathcal{C}^{(q)}$ over $\Bbb F_q$, generated by the matrix obtained by replacing each entry of $G$ with its column representation (a column vector in $\Bbb F_q^m$) with respect to $\mathcal{B}$, is termed a subfield code. In \cite{WLX,SS4,SS,SS5,LY}, many codes exhibit outstanding parameters in their subfield codes.

\subsection{The Projective Solomon-Stiffler Codes}

A linear code $\mathcal{C}$ is called projective if its dual code has minimum distance at least $3$. Thus the dual of a projective code has better error correcting capability compared to that of a nonprojective code.

We recall the concept of projective space to introduce the projective Solomon-Stiffler code. The projective space (also called projective geometry) PG$(m-1,q)$ is the set of all the 1-dimensional subspaces (called points) of the vector space $\Bbb F_q^m$. For any $r$-dimensional $\Bbb F_q$-subspace $H$ of $\Bbb F_q^m$ , the set of all the $1$-dimensional subspaces of $H$ is said to be a projective subspace of PG$(m-1,q)$; this projective subspace has the projective dimension $r - 1$ which is one less than the dimension of the corresponding subspace $H$.
Let $m\geq2$. A point of PG$(m-1,q)$ is given in homogeneous coordinates by $\mathbf{x}=(x_{0},x_{1},\ldots,x_{m-1})\in\mathbb{F}_{q}^{m}$ where all $x_{i}$’s are not all zero (which implies $\mathbf{0}\notin\mathrm{PG}(m-1,q)$); each point has $q-1$ coordinate representations, since ${\bf x}$ and $\lambda{\bf x}$ yield the same $1$-dimensional subspace of $\Bbb F_q^m$ for any $\lambda\in\mathbb{F}_{q}^{*}$.

By definition, it can be readily verified that the code $\mathcal{C}_{D}$ defined by \eqref{eq2.1} is projective if and only if any two elements of $D$ are linearly independent over $\Bbb F_q$.
\begin{equation*}
\mathbb{F}_{q^m}^*=\mathbb{F}_q^*\overline{\mathbb{F}_{q^m}^*}=\{yz:y\in\mathbb{F}_q^* ~\mbox{ and }~  z\in\overline{\mathbb{F}_{q^m}^*}\},
\end{equation*}
where $z_{i}/z_{j}\notin\mathbb{F}_{q}^{*}$ for distinct elements $z_{i}$  and $z_{j}$ in $\overline{\mathbb{F}_{q^m}^*}$. The set of all points in PG$(m-1,q)$ can be identified with the set $\overline{\mathbb{F}_{q^m}^*}$ since $\mathbb{F}_{q}^{m}$ is identical to $\mathbb{F}_{q^m}$. Therefore, the code $\mathcal{C}_D$ defined by \eqref{eq2.1} is projective if $D \subseteq\mathrm{PG}(m-1,q)$.

\subsection{Two Bounds of Linear Codes}

\begin{lemma}{\rm (Griesmer Bound \cite{G}) For a given  $[n, k, d]$  linear code over $\Bbb F_q$,  there is a bound 
$$\sum_{i=0}^{k-1}\bigg\lceil {\frac{d}{q^i}} \bigg\rceil \le n,$$ 
where $\lceil {x} \rceil$ denotes the smallest integer greater than or equal to $x$.

}
\end{lemma}

A linear code $\mathcal{C}$  with parameters $[n, k, d]$ is distance optimal if there is no $[n,k,d+1]$ code. Conversely, if the code $[n,k,d+1]$ is optimal, then $\mathcal{C}$ is referred to as almost optimal (see \cite[Chapter 2]{HP}). When a linear code meets the Griesmer bound, it is classified as a Griesmer code. It can be verified that Griesmer codes are distance optimal.

\begin{lemma}{\rm  (Sphere Packing Bound \cite{HP})   
For a given  $[n, k, d]$  linear code over $\Bbb F_q$,  there is a bound as follows:
 $$\sum_{i=0}^{\left \lfloor \frac{d-1}{2} \right \rfloor} \dbinom{n}{i} (q-1)^{i}\le q^{n-k}, $$
  where $ \left \lfloor {\cdot}  \right \rfloor$ is the floor function. 
   }
 \end{lemma}

When a code achieves the bound mentioned above with equality, it is referred to as a {\it perfect code}. It can be verified that perfect codes are also distance optimal.

\subsection{Simplicial Complexes and Posets}

Let $\Bbb F_2$ denote the finite field with two elements, and let $m$ be a positive integer. The support of a vector ${\bf v} \in \Bbb F_2^m$ is defined as the set of nonzero coordinates and is denoted by $\mathrm{supp}({\bf v})$. The Hamming weight $\mbox{wt}({\bf v})$ of ${\bf v}$ in $\mathbb{F}^m_2$ is defined as the size of its support. For two subsets $A$ and $B$ of $[m]$, the set $\{x : x \in A \text{ and } x \notin B\}$ is denoted by $A\backslash B$, and the number of elements in set $A$ is denoted by $|A|$.

Given two vectors ${\bf u}$ and ${\bf v}$ in $\mathbb{F}_2^m$, we say that ${\bf v}$ is a subset of ${\bf u}$, denoted by ${\bf v} \subseteq {\bf u}$, if the support of ${\bf v}$ is a subset of the support of ${\bf u}$. A family $\Delta \subseteq \mathbb{F}_2^m$ is termed a simplicial complex if it contains every subset ${\bf u}$ whenever a subset ${\bf v}$ is a subset of ${\bf u}$ for all ${\bf v}$ in $\Delta$. The maximal elements of $\Delta$ are those that are not properly contained in any other element of $\Delta$. Let $\mathcal{F} = \{F_1, \ldots, F_l\}$ denote the family of maximal elements of $\Delta$. For each $F \subseteq [m]$, the simplicial complex $\Delta_F$, generated by $F$, is defined as the family of all subsets of $F$.

Consider a simplicial complex $\Delta \subseteq \mathbb{Z}_2^n$. Hyun et al. introduced the following $n$-variable generating function associated with the set $\Delta$:
$$ \mathcal{H}_{\Delta}(x_1,x_2,\ldots, x_n)=\sum_{\bv\in \Delta}\prod_{i=1}^nx_i^{v_i}\in \mathbb{Z}[x_1,x_2, \ldots, x_n], $$
where ${\bf v} = (v_1,v_2,\ldots ,v_n) \in \mathbb{Z}_2^n$.

A partially ordered set (abbreviated as poset) $\mathbb{P} = ([n],\preceq)$ is defined as a partial order relation on $[n]$, satisfying reflexivity, antisymmetry, and transitivity properties for all $i,j,k \in [n]$. Two distinct elements $i$ and $j$ in $[n]$ are considered comparable if either $i\preceq j$ or $j\preceq i$, and otherwise, they are incomparable. If every pair of distinct elements in a poset $\mathbb{P}$ is incomparable, then $\mathbb{P}$ is termed an antichain. An order ideal is a nonempty subset of $\mathbb{P}$ such that if an element $j$ is in the subset and $i \preceq j$, then $i$ must also be in the subset. The smallest order ideal of $\mathbb{P}$ containing a subset $E$ is denoted by $\langle E\rangle$. For an order ideal $I$ of $\mathbb{P}$, $I(\mathbb{P})$ denotes the set of order ideals of $\mathbb{P}$ contained in $I$. Let $\mathcal{I}=\{I_1,\ldots, I_m\}$ be a subset of $\mathcal{O}_{\mathbb{P}}$. We define
\begin{equation}
\mathcal{I}(\mathbb{P}) = \{J\in \mathcal{O}_{\mathbb{P}}: J\subseteq I\in \mathcal{I}\} = \bigcup_{i=1}^m I_i(\mathbb{P}).
\end{equation}
Then $\mathcal{I}(\mathbb{P})$ forms an order ideal of $\mathcal{O}_{\mathbb{P}}$ with the partial order $\subseteq$.
It is worth noting that if $\mathbb{P}$ is an antichain, then $\mathcal{I}(\mathbb{P})$ constitutes a simplicial complex.



In \cite{H23}, Hu {\em et al.}, in a more generalized approach, defined $A_{i}=\mathrm{supp}(F_{i})$ for $1\leq i\leq h$, implying $A_{i}\subseteq[m]$. It's noted that $A_i\setminus A_j\neq\emptyset $ for any $1 \leq i\neq j\leq h$ by definition. Let $\mathcal{A}=\{A_{1},A_{2},\ldots,A_{h}\}$ represent the set of supports of maximal elements of $\Delta$, and $\mathcal{A}$ is termed the support of $\Delta$, denoted supp($\Delta$) = $\mathcal{A}$. Then one can observe that a simplicial complex $\Delta$ is uniquely generated by $\mathcal{A}$, denoted $\Delta=\langle\mathcal{A}\rangle $. 

For any set $\mathcal{K}$ consisting of some subsets of $[m]$, it is stated that a simplicial complex $\Delta$ of $\mathbb{F}_{q}^{m}$ is generated by $\mathcal{K}$, denoted $\Delta=\langle\mathcal{K}\rangle $, if $\Delta$ is the smallest simplicial complex of $\mathbb{F}_{q}^{m}$ containing every element in $\mathbb{F}_{q}^{m}$ with the support $K\in\mathcal{K}$.
Consider a simplicial complex $\Delta \subseteq \mathbb{F}_2^n$ and $\Delta^{c}$ being the complement of $\Delta$. 
Hence for any simplicial complex $\Delta$, $\Delta^{c}$ can be expressed as
\begin{equation*}
\Delta^{c}=\mathbb{F}_{q}^{*}\overline{\Delta}^{c}=\{yz:y\in\mathbb{F}_{q}^{*}\mathrm{~and~}z\in\overline{\Delta}^{c}\},
\end{equation*}
where $z_{i}/z_{j}\notin\mathbb{F}_{q}^{*}$ for distinct elements $z_{i}$  and $z_{j}$ in $\overline{\Delta}^{c}$, and clearly $|\overline{\Delta}^{c}|=|\dot{\Delta}^{c}|/(q-1)$. This defines ${\Delta}^{c}$ and $\overline{\Delta}^{c}$ can be viewed as a subset of $PG(m-1,q)$.

\section{Codes over finite rings}

In this section, we will discuss the codes over finite rings.

\subsection{Codes over ring  $\mathbb Z_4$}

 
 Wu {\em et al.} \cite{WLZ} found some new quaternary codes by using simplicial complexes  and the following results are documented.

\begin{theorem} {\rm  \cite[Theorem 3.2]{WLZ}\label{thm1-1} 
Let $n \geq 2$ be an integer and $A,B \subseteq [n]$.
Assume that $\Delta$ is a simplicial complex of $\mathbb{F}_2^n$ with the two maximal elements $A,B$ and  $\Delta_A$ is a simplicial complex of $\mathbb{F}_2^n$. 
 Let $D = \Delta_A + 2\Delta^*$. Then the code  $\mathcal C_{D}$ in \eqref{eq2.1} has length $|D|=2^{|A|}(2^{|A|}+2^{|B|}-2^{|A \cap B|}-1)$ and size $4^{|A|}2^{|B\backslash  A| }$. Its Lee weight distribution is given in \cite[Table I]{WLZ}.
}
\end{theorem}

\begin{theorem}{\rm\cite[Theorem 4.4]{WLZ}   In Theorem \ref{thm1-1},   the Gray image $\phi (\mathcal{C}_{D})$ is linear  if and only if $|A|=1$.

}

\end{theorem}

In \cite{ZWY}, Zhu {\em et al.} found that  the code $\mathcal {C}_{D}$ in \eqref{eq2.1} could obtain some new quaternary codes comparing with $\Bbb Z_4$-database as shown in Table \ref{tabZWYnewcode} by the posets of disjoint union of two chains.

For $m\in [n]$, the posets of disjoint union of two chains, denoted by $\mathbb{P}=(m\oplus n, \preceq)$ and its  Hasse diagram is given  in Figure 1. Let $I$ be an order ideal in $\mathbb{P} =(m \oplus n, \preceq) $, $L = (I(\Bbb P))^c = \Bbb F_2^n \backslash I(\Bbb P)$, the defining set $D = L + 2\Bbb F_2^n \subset \Bbb Z_4^n$.

\[
\begin{tikzpicture}
\draw (0, -1)[fill = black] circle (0.05);
\draw (0, -2)[fill = black] circle (0.05);
\draw (0, -2.5)[fill = black] circle (0.05);
\draw (0, -3)[fill = black] circle (0.05);
\draw (0, -3.5)[fill = black] circle (0.05);
\draw (0, -4)[fill = black] circle (0.05);
\draw (0, -5)[fill = black] circle (0.05);

\draw (3, -1)[fill = black] circle (0.05);
\draw (3, -2)[fill = black] circle (0.05);
\draw (3, -2.5)[fill = black] circle (0.05);
\draw (3, -3)[fill = black] circle (0.05);
\draw (3, -3.5)[fill = black] circle (0.05);
\draw (3, -4)[fill = black] circle (0.05);
\draw (3, -5)[fill = black] circle (0.05);

\draw[thin] (0, -5) --(0,-4);

\draw[thin] (0,-2)--(0,-1);
\draw[thin] (3, -5) --(3,-4);

\draw[thin] (3,-2)--(3,-1);

\node  at (0, -1) [left] {$~~~m~~$};
\node  at (0, -2) [left] {$~~~m-1~~$};
\node  at (0, -4) [left] {$~~~2~~$};
\node  at (0, -5) [left] {$~~~1~~$};
\node  at (3, -4) [right] {$~~m+2~~$};
\node  at (3, -5) [right] {$~~m+1~~$};
\node  at (3, -1) [right] {$~~~n~~$};\node  at (3, -2) [right] {$~~~n-1~~$};

\node  at (2, -5.5) [below] {Figure 1 $\mathbb{P}=$($m\oplus n, \preceq$)};

\end{tikzpicture}
\]

\begin{table}[h]  
\begin{center}
\caption{Linear quaternary codes $\mathcal {C}_{D}$ in \cite{ZWY}}   
\begin{tabu} to 0.6\textwidth{X[0.6,c]|X[2,c]}  
\hline 
\rm{Parameters}&\rm{Remark}\\ 
\hline
$(4,8,4)_L$&New\\ 
\hline
$(32,64,16)_L$& New \\ 
\hline
$(32,64,32)_L$& New (Only a nonlinear code with same parameters) \\ 
\hline
$(48,64,48)_L$& New (Only a nonlinear code with same parameters) \\ 
\hline
\end{tabu}
\label{tabZWYnewcode}  
\end{center}
\end{table}

From the above results, it would be natural to consider the following problems.

\begin{problem} {\rm  Find more codes over $\mathbb Z_4$ with few Lee weight and their Gray image possessing some special properties  by using simplicial complexes or posets. 

}
\end{problem}

\begin{problem} {\rm  Find more new codes over $\mathbb Z_4$   by using simplicial complexes or posets. 

}
\end{problem}

\subsection{Codes over the ring $\Bbb F_p + u\Bbb F_p$ with $u^2=0$ }

For $R=\Bbb F_p + u\Bbb F_p$, where $u^2 = 0$, let $R_m$ be an extension of $R$ of degree $m$, such that the vector space is the same as $V_m$ in \eqref{eq2.1}. The code $\mathcal {C}_{D}$ with the defining set $D = \Delta_1+u\Delta_2 \subseteq R^m$ was discussed in \cite{WZY} for $p=2$ and in \cite{WH} for odd prime $p$.

The Gray map $\hat{\phi}$ from $R$ to $\Bbb F_p^2$ is defined by
$$\hat{\phi}: R\to \Bbb F_p^2, a+ub\mapsto (b,a+b),~a,b\in \Bbb F_p.$$
This naturally leads to the Gray map $\phi$ from $R^m$ to $\Bbb F_p^{2m}$ as follows:
$$\phi: R^m\to \Bbb F_p^{2m},~\mathbf{x}=\mathbf{a}+u\mathbf{b}\mapsto (\mathbf{b}, \mathbf{a}+\mathbf{b}).$$

For the codes over $R=\Bbb F_p + u\Bbb F_p$ with $u^2 = 0$, the following results are documented.

\begin{theorem}{\rm \cite[Theorems 3.2 and 3.6]{WZY}\label{WZYthm3.2} ~ Let $\Delta_A, \Delta_B$ be two simplicial complexes of $\Bbb F_2^m$ with two maximal elements $A,B$, respectively.

$(1)$ Let $D_1=\Delta_A+u\Delta_B^c$, then the code $\mathcal{C}_{D_1}$ in \eqref{eq2.1} has length $2^{|A|}(2^{m}-2^{|B|})$, size $2^{m+|A|}$, and its Lee weight distribution is given in \cite[Table 1]{WZY}.

$(2)$ Let $D_2=\Delta_A^c+u\Delta_B^c$, then the code $\mathcal{C}_{D_2}$ in \eqref{eq2.1} has length $(2^{m}-2^{|A|})(2^{m}-2^{|B|})$, size $2^{2m}$, and its Lee weight distribution is given in \cite[Table 4]{WZY}.

}
\end{theorem}

\begin{theorem}{\rm\cite[Theorems 4.1 and 4.2]{WZY}\label{WZYthm4.1}~ Keeping the notion in Theorem \ref{WZYthm3.2}.

$(1)$ The binary code $\phi(\mathcal{C}_{D_1})$ with parameters $[2^{|A|+1}(2^m-2^{|B|}),m+|A|,2^{|A|}(2^m-2^{|B|})]$ in Theorem \ref{WZYthm3.2} is distance optimal.

$(2)$ If $A=B$ and $|A|=m-1$, then the binary code $\phi(\mathcal{C}_{D_2})$ has parameters $[2^{2m-1}, 2m, 2^{2m-2}]$ meeting the Griesmer bound with equality.

}
\end{theorem}

\begin{theorem}{\rm \cite[Theorems 4.1 and 4.2]{WH} 
Let $m\ge 3$ be an integer and $p$ be an odd prime number.
Let $L=\Delta^c+u\mathbb{F}^m_p,u^2=0$.

$(1)$ If $\Delta=\langle(r,0,\ldots,0)\rangle$, $r=1,2,\ldots,p-1$, then the Gray image $\phi(\mathcal{C}_{L})$ of $\mathcal{C}_{L}$ is a distance optimal linear code, and the minimum distance of $\phi(\mathcal{C}_{L})^{\perp}$ is two.
In particular, if $p<2r+2$ and $2(r+1)(p-1)<p^2$, i.e., $r=
\frac{p-1}{2}$, then $\phi(\mathcal{C}_{L})$ meets the Griesmer bound.

$(2)$ If $\Delta=\langle(p-1,r,0,\ldots,0)\rangle$, $r=1,2,\ldots,p-1$, then the Gray image $\phi(\mathcal{C}_{L})$ of $\mathcal{C}_{L}$ is a distance optimal linear code, and the minimum distance of $\phi(\mathcal{C}_{L})^{\perp}$ is two.
In particular, if $p<2r+2$ and $2(r+1)(p-1)<p^2$, i.e., $r=
\frac{p-1}{2}$, then $\phi(\mathcal{C}_{L})$ meets the Griesmer bound.
}
\end{theorem}

For the ring $R=\Bbb F_2 + u\Bbb F_2 + u^2 \Bbb F_2$, where $u^3 = 0$, any $a+ub+u^2c\in R$, $a,b,c\in \Bbb F_2$, the Gray map $\phi$ from $R$ to $\Bbb F_2^3$ is defined by 
$$\phi: R\to \Bbb F_2^3, a+ub+u^2c\mapsto (c,b+c,a+b+c).$$
This Gray map $\phi$ can be extended naturally from $R^m$ to $\Bbb F_2^{3m}$.

For the defining set $D = \Delta_1 + u\Delta_2 +u^2\Delta_3$, the code $\mathcal{C}_D$ in \eqref{eq2.1} when the simplicial complexes are generated by a single maximal element was discussed in \cite[Theorem 4]{LS}. Moreover, the binary codes obtained under the Gray image are minimal and optimal.

\begin{theorem}{\rm\cite[Corollary 6]{LS}
~Let $\Delta_A, \Delta_B$ and $\Delta_C$ be three simplicial complexes of $\Bbb F_2^m$, and $D=\Delta_A + u\Delta_B^c+ u^2\Delta_C$.
When $A=C=[m]$ and $|B| = m-1$, the code $\phi(\mathcal{C}_D)$ is a three-weight binary code with parameters $[3\times2^{3m-2},3m,3\times2^{3m-2}]$, which is minimal and distance optimal. Its weight distribution is as follows: $A_0 = 1$, $ A_{3\cdot2^{3m-2}} = 2^{3m} -3 $ and $A_{2^{3m}} = A_{5\cdot2^{3m-2}} = 1$.

}
\end{theorem}

The ring $R = \Bbb F_2 + u\Bbb F_2 + \cdots + u^{k-1}\Bbb F_2$, where $u^k = 0$, can be naturally extended to degree $m$, which is the same as $V_m$. Let $D = \Delta_1 + u\Delta_2 + \cdots + u^{k-1}\Delta_k \subseteq \mathcal{R}$ be the defining set. The code $\mathcal{C}_D$ in \eqref{eq2.1} with two different sets $D$ was studied in \cite[Theorems 1 and 2]{LS2}. Moreover, the binary codes obtained from these two codes are distance optimal.

Any integer $z$ can be written uniquely in base $p$ as $z=p_0(z)+pp_1(z)+p^2p_2(z)+\cdots$, where $p$ is prime and $0\leq p_i(z) \leq p-1$, $i=0,1,2,\cdots$.
For  any $a=a_0+a_1u+\cdots+a_{k-1}u^{k-1}\in R$, $a_i \in \Bbb F_2$, the Gray map $\phi$ from $R$ to $\Bbb F_2^{2^{k-1}}$ is defined by 
$$\phi: R\to \Bbb F_2^{2^{k-1}}, a \mapsto (b_0,b_1,\cdots,b_{2^{k-1}-1}),$$
where all $0\leq i \leq 2^{k-2}-1$, $0\leq \varepsilon \leq 1$ and $b_{2i+\varepsilon} = a_{k-1}+\sum_{l=1}^{k-2}p_{l-1}(i)a_l+\varepsilon a_0$.
This Gray map $\phi$ can be extended naturally from $R^m$ to $\Bbb F_2^{2^{k-1}}$.

\begin{theorem}{\rm\cite[Propositions 1-3]{LS2}\label{LS2pro1}
 For an integer $k \geq 3$, the following two codes are minimal with respect to the Ashikhmin–Barg condition.

$(1)$ Let $D_1 = \Bbb F_2^m + u\Bbb F_2^m + \cdots + u^{k-1}\Delta_A$ with certain $|A| = m$. Then the binary code $\phi(\mathcal{C}_{D_1})$ has parameters $[2^{k(m+1)-1},km,2^{k(m+1)-2}]$, which is distance optimal.

$(2)$ Let $D_2 = \Bbb F_2^m + u\Bbb F_2^m + \cdots + u^{k-1}\Delta_A^c$. If certain $|A|$ satisfies one of the following conditions,

\quad $(i)$ $2^{k-1}+m-km-k < |A| \leq m-k+1,$

\quad $(ii)$ $S < |A| \leq m-1$,

where $S = \max\{ m-k+1, \log_2(2^{k-1}-km)+m 1-k \}$, then the binary code $\phi(\mathcal{C}_{D_2})$ has parameters $[2^{(k-1)(m+1)}(2^m-2^{|A|}),km,2^{(k-1)(m+1)-1}(2^m-2^{|A|})]$, which is distance optimal.

}
\end{theorem}


\subsection{Codes over ring $\Bbb F_p + u\Bbb F_p$ with $u^2=u$ }

The ring $R = \Bbb F_p + u\Bbb F_p$, where $u^2 = u$, can naturally extend to degree $m$, which is the same as $V_m$ in \eqref{eq2.1}. Then, the code $\mathcal {C}_{D}$ in \eqref{eq2.1} with the defining set $D = u\Delta_1 + (1-u)\Delta_2 \subseteq R^m$ was discussed in \cite{SLP}. 
For the ring $R=\Bbb F_p + u\Bbb F_p$, where $u^2 = u$, any $a+ub\in R$, $a,b\in \Bbb F_p$, the Gray map $\phi$ from $R$ to $\Bbb F_p^2$ is defined by 
$$\phi: R\to \Bbb F_p^2, a+ub\mapsto (-b,2a+b).$$
This Gray map $\phi$ can be extended naturally from $R^m$ to $\Bbb F_p^{2m}$.

For the codes over $R=\Bbb F_p + u\Bbb F_p$ with $u^2 = u$, the following results are documented.

\begin{theorem}{\rm\cite[Theorems 1-2]{SLP}\label{SLPthm1} ~Let $\Delta_1,\Delta_2$ be two down-sets of $\Bbb F_q$, $\bv, \bv^* \in \Bbb F_p^m$, $A = \mathrm{supp}(\bv)$, and $B=\mathrm{supp}(\bv^*)$ such that $v_i=v_j^*=p-1$, where $i\in A$ and $j \in B$. Let $\Delta_1=\langle \bv\rangle$, $\Delta_2=\langle \bv^*\rangle$.

$(1)$ Let $D_1=u\Delta_1+(1-u)\Delta_2^c$. Then the code $\mathcal {C}_{D_1}$ in \eqref{eq2.1} has length $p^{|A|}(p^{m}-p^{|B|})$, size $p^{2m}$, and its Lee weight distribution is given in \cite[Table 1]{SLP}.

$(2)$ Let $D_2=u\Delta_1^c+(1-u)\Delta_2^c$. Then the code $\mathcal {C}_{D_2}$ in \eqref{eq2.1} has length $(p^{m}-p^{|A|})(p^{m}-p^{|B|})$, size $p^{2m}$, and its Lee weight distribution is given in \cite[Table 2]{SLP}.

}

\end{theorem}

\begin{theorem}{\rm\cite[Propositions 1-2]{SLP}\label{SLPpro1} ~ Keep the notion in Theorem \ref{SLPthm1}.

$(1)$ If $|A|+|B|\geq 4$, then the binary code $\phi(\mathcal{C}_{D_1})$ has parameters $[2p^{|A|}(p^{m}-p^{|B|}),m+|A|,2(p-1)p^{|A|-1}(p^{m}-p^{|B|})]$ in Theorem \ref{SLPthm1}$(1)$ and is distance optimal.

$(2)$ If $|A|=|B|= m-1$ and $p=2$, then the binary code $\phi(\mathcal{C}_{D_2})$ has parameters $[2^{2m-1},2m-1,2^{2m-2}]$ in Theorem \ref{SLPthm1}$(2)$ and is distance optimal.

}

\end{theorem}



\subsection{Codes over rings $\Bbb F_2 + u\Bbb F_2 + v\Bbb F_2 + uv \Bbb F_2$, where $u^2 = 0$, $v^2 = 0$, and $uv = vu$}

Let $R=\Bbb F_2 + u\Bbb F_2 + v\Bbb F_2 + uv \Bbb F_2$, where $u^2 = 0$, $v^2 = 0$, and $uv = vu$. For the defining set $D = \Delta_1 + u\Delta_2 +v\Delta_3 + uv\Delta_4 \subseteq R^m$, the code $\mathcal{C}_D$ in \eqref{eq2.1} with five different defining sets $D$ was discussed in \cite[Theorems 1-5]{SL2}.

For  any $a+bu+cv+duv\in R$, $a,b,c,d\in \Bbb F_2$, the Gray map $\phi$ from $R$ to $\Bbb F_2^4$ is defined by 
$$\phi: R\to \Bbb F_2^4, a+bu+cv+duv\mapsto (d,c+d,b+d,a+b+c+d).$$
This Gray map $\phi$ can be extended naturally from $R^m$ to $\Bbb F_2^{4m}$.

\begin{theorem}{\rm\cite[Proposition 1]{SL2}\label{SL2pro1}
Let $\Delta_A$, $\Delta_B$, and $\Delta_C$ be three simplicial complexes of $\Bbb F_2^m$. All below three binary codes are minimal if $|A| \geq 3$ or $|{C}| \leq m-1$ with respect to the Ashikhmin-Barg condition.

$(1)$ Let $D_1=uv\Delta_A$, the Lee weight distribution of code $\mathcal{C}_{D_1}$ is given in \cite[Table 1]{SL2}. Then $\phi(\mathcal{C}_{D_1})$ is a 1-weight linear code with parameters $[2^{|{A}|+2},|{A}|,2^{|\text{A}|+1}]$, which is distance optimal.

$(2)$ Let $D_2=uv\Delta_A^c$, the Lee weight distribution of the code $\mathcal{C}_{D_2}$ is given in \cite[Table 2]{SL2}. Then $\phi(\mathcal{C}_{D_2})$ is a 2-weight linear code with parameters $[2^{m+1}-2^{|{A}|+1},4m,2^{m+1}-2^{|\text{A}|+1}]$, which is distance optimal if $|\text{A}|\geq 3$.

$(3)$ Let $D_3=u\Bbb F_2^m + v\Delta_B + uv\Delta_C^c$, the Lee weight distribution of code $\mathcal{C}_{D_3}$ is given in \cite[Table 7]{SL2}. Then $\phi(\mathcal{C}_{D_3})$ is a 3-weight linear code with parameters $[2^{m+|\text{B}|+1}(2^m-2^{|{C}|}),4m,2^{m+|\text{B}|+1}(2^m-2^{|C|})]$, which is distance optimal if $m \geq 2$.

}

\end{theorem}

\subsection{$\Bbb Z_p\Bbb Z_p[u]$-Additive codes}

Let $\Bbb Z_p[u]=\Bbb Z_p+u\Bbb Z_p $, where $u^2=u$ and  $p$ is an odd prime. Define $$\mathcal{R}=\Bbb Z_p\Bbb Z_p[u]={ (\lambda,\mu+u\nu)|\lambda,\mu,\nu \in \Bbb Z_p }.$$ Set defining set $D =\{ {\bf t}_1, u{\bf t}_2+(1-u){\bf t}_3 | {\bf t}_1 \in \Delta_1, {\bf t}_2 \in \Delta_2, {\bf t}_ 3\in \Delta_3\}$ then the code $\mathcal{C}_{D}$ in \eqref{eq2.1} associated with down-sets was introduced in \cite[Theorem 3.4]{WS}.

Any $\mu+u\nu\in \Bbb Z_p[u]$, $\mu,\nu\in \Bbb Z_p$, the Gray map $\phi^{\prime}$ from $\Bbb Z_p[u]$ to $\Bbb Z_p^2$ is defined as 
$$\phi^{\prime}: \Bbb Z_p[u]\to \Bbb Z_p^2, \mu+u\nu \mapsto (-\mu,2\mu+\nu).$$
Any $\mu+uv\in \mathcal{R}$, $\lambda,\mu,\nu\in \Bbb Z_p$, the Gray map $\phi$ from $\mathcal{R}$ to $\Bbb Z_p^3$ is defined as 
$$\phi: \mathcal{R}\to \Bbb Z_p^3, (\lambda,\mu+u\nu) \mapsto (\lambda,-\mu,2\mu+\nu).$$
This Gray map $\phi$ can be extended naturally from $\mathcal{R}^m$ to $\Bbb Z_p^{3m}$.

\begin{theorem}{\rm\cite[Proposition 3.6]{WS}\label{WSpro3.6}
~Let $\Delta_A, \Delta_B, \Delta_C \subseteq \Bbb Z_p^m$ be three down-sets of $\Bbb F_2^m$ with three maximal elements $A,B,C$, respectively. Let $D =\{ {\bf t}_1, u{\bf t}_2+(1-u){\bf t}_3 | {\bf t}_1 \in \Delta_A^c, {\bf t}_2 \in \Delta_B^c, {\bf t}_3 \in \Delta_C^c\}$. Then the code $\mathcal{C}_{D}$ in \eqref{eq2.1} has length $(p^{m}-p^{|A|})(p^{m}-p^{|B|})(p^{m}-p^{|C|})$, size $p^{3m}$ and the Lee weight distribution is shown in \cite[Table 1]{WS} .
The code $\phi(\mathcal{C}_{D})$ is distance optimal when $|A| + |B| + |C| \geq 2$. 

}
\end{theorem}

\subsection{Codes over a special ring}

The commutative and non-unital ring $I$ is defined as $I = \langle a, b | 2a = 2b = 0, a^2 = b, b^2 = b, ab = 0 \rangle$ and the underlying sets of $I=\{0,a,b,c\}$. $I^m$ is an extension of $I$ degree $m$, which can be represented as $I^m = a\Bbb F_2^m + b\Bbb F_2^m$.

The code $C_D$ in \eqref{eq2.1} over $I$ with defining set $D= a\Delta_1+b\Delta_2 \subseteq I^m$ in four cases was introduced in \cite[Theorem 3.1]{SS3}. Notably, one class of binary code obtained from the $C_D$ under the Gray image is minimal and optimal.

The Gray map $\hat\phi$ from $I$ to $\Bbb F_p^2$ is defined by 
$$\hat{\phi}: I\to \Bbb F_2^2, as+bt\mapsto (t,s+t),~s,t\in \Bbb F_2.$$
This leads to the Gray map $\phi$ naturally from $I^m$ to $\Bbb F_2^{2m}$ as :
$$\phi: I^m \to \Bbb F_2^{2m}, \mathbf{x}=a\mathbf{s}+b\mathbf{t} \mapsto (\mathbf{t},\mathbf{s+t}).$$

\begin{theorem}{\rm\cite[Theorem 4.3]{SS3}\label{SS3thm4.3.2}
~Let $D = a\Delta_M^c + b\Delta_N \subseteq I^m$. Then the $\phi(C_D)$ is a 2-weight linear binary code with parameters $[(2^m-2^{|M|})2^{|N|+1}, m, (2^m-2^{|M|})2^{|N|}]$.

$(1)$ Let $1 \leq |M| + |N|\leq m-1$. If $0 < 2^{|N|+ 1}-1 < |M| + |N|$ then $\phi(C_D)$ is optimal with respect to the Griesmer bound.

$(2)$ Let $m \leq |M| + |N| \leq 2m-1$. If $0 < 2^{|M|+|N|+1-m}(2^{m-|M|}- 1) < m$ then $\phi(C_D)$ is optimal with respect to the Griesmer bound.

}
\end{theorem}

\subsection{Codes over mixed alphabet ring}

We list some codes over some special finite rings constructed from simplicial complexes.

\begin{itemize}
\item Let $R = \Bbb Z_2 [u] \backslash \langle u^3 - u \rangle$ such that $R$ can be written as $(1+u^2)\Bbb Z_2+u^2(\Bbb Z_2 + (u+u^2)\Bbb Z_2)$ uniquely. $R^m$ is an extension of $R$ of degree $m$, which can be considered the same as $V_m$ in \eqref{eq2.1}. Then the code $\mathcal{C}_D$ with defining set $D = (1+u^2)\Delta_1+u^2(\Delta_2 + (u+u^2)\Delta_3)$ in \eqref{eq2.1} over such $R$ is discussed in \cite[Theorems 3.4]{SS2} when choosing different defining sets $D$ in 8 cases. Furthermore, by applying the Gray image, the authors obtained binary codes in \cite[Theorem 4.4]{SS2}, and among them, certain codes were found to be self-orthogonal and minimal with respect to the condition of Ashikhmin-Barg.

\item Let $\Bbb Z_2[u]=\Bbb Z_2 +u\Bbb Z_2$, $u^2=0$. Define $R=\Bbb Z_2\Bbb Z_2[u]=\{ (x,y+uz)|x,y,z \in \Bbb Z_2 \}$ and $R^m$ as an extension of $R$ of degree $m$, $\mathcal{R}=R^m =\Bbb Z_2\Bbb Z_2[u]=\{ (p,q+ur)|p,q,r \in \Bbb Z_2^m \}$. Then the code $\mathcal{C}_D$ in \eqref{eq2.1} with defining set $D = \{(t_1,t_2+ut_3)|t_1\in \Delta_1, t_2\in \Delta_2,t_3\in \Delta_3\}$ is introduced in \cite[Theorem 3.2]{KM}, which answers the question raised in \cite{LY} about the use of simplicial complexes or down-sets in the case of mixed alphabet rings. By applying the Gray image, the binary code is self-orthogonal in \cite[Proposition 4.1]{KM} and minimal in \cite[Theorem 4.5]{KM} for specific conditions.

\item The non-commutative and non-unital ring $E$ is defined as $E = \langle a, b~ | ~2a = 2b = 0, a^2 = a, b^2 = b, ab = a, ba = b \rangle$ and the underlying sets of $E=\{0,a,b,c=a+b\}$. $E^m$ is an extension of $E$ of degree $m$, which can be considered as $E^m = a\Bbb F_2^m+c\Bbb F_2^m$. Then the left-$E$-codes $C_D^L$ with four cases defining set $D = a\Delta_1+c\Delta_2 \subseteq E^m$ in \cite[Theorem 3.1]{SS4} are introduced. In the construction of codes $C_D^L$, there have not been remarkably outstanding parameters. However, it is worth mentioning that the binary codes obtained through the Gray map are self-orthogonal in \cite[Proposition 3.3]{SS4}. By calculating the subfield-like codes  with respect to the $C_D^L$, a class of optimal binary code is obtained.
\end{itemize}

For codes over finite rings, we also have the following problem:

\begin{problem}{\rm Find more optimal codes over different finite rings and determine their weight distributions in different metrics. 

}
\end{problem}

\section{Codes over finite fields}

In this section, we will discuss some codes over finite fields which are all related with simplicial complexes.

\subsection{Binary codes associate with Boolean function}

 Chang and Hyun \cite{CH} introduced a new method for constructing linear codes by leveraging simplicial complexes. They also introduce $n$-variable generating functions and present Walsh–Hadamard transforms related to certain simplicial complexes.


\begin{theorem}{\rm\cite[Propositions 1-3]{CH}\label{CHpro1}  Let $\Delta$ be a simplicial complex of $\Bbb Z_2^n$ and $f = f_{\Delta^*}$ be the Boolean function in $n$-variable whose support is $\Delta^*$.

(1) If $\Delta$ has only one maximal element $A$, then the code $\mathcal{C}_{f}$ in \eqref{eq2.2} has parameters $[2^n-1, n+1]$ and its Hamming weight distribution is given. Moreover, it is a two-weight linear code with parameters $[2^n -1,n+1,2^{n-1}-1]$ when $|A|=n$. Thus there exist non-equivalent optimal linear codes with parameters $[2^n-1,n+1,2^{n-1}-1]$ that attain the Griesmer bound with equality.

(2) If $\Delta$ has two maximal elements $A, B$, then the code $\mathcal{C}_{f}$ in \eqref{eq2.2} is minimal violating the condition of Ashikhmin-Barg if and only if $ A \cup B=[n] $ and max$\{|A|,|B|\}\leq n-2$.

}
\end{theorem}






%

After that, some results are extended to posets as follows:

\begin{theorem}{\rm\cite[Theorems 5.5 and 6.3]{HKWY}\label{HKWYthm5.5}
Let $\Bbb H(m,n)$ be a hierarchical poset with two levels. Let $I=A \cup B$ be an order ideal of $\Bbb H(m,n)$, where $A\subseteq [m]$, $B\subseteq [n]\backslash[m]$. Set $\mathcal{I}={I}$.

$(1)$ If $B=\emptyset$, then the weight distribution of the code $\mathcal{C}_f$ is given in \cite[Table 5]{HKWY}.

$(2)$ If $B\neq\emptyset$, then the weight distribution of the code $\mathcal{C}_f$ is given in \cite[Table 6]{HKWY}.
For the subcase $(2)$ note that:

\quad $(i)$ If $m=n-1$ and $|B| =1$, then the code $\mathcal{C}_f$ is an almost optimal linear code with parameters $[2^n-1,n+1,2^{n-1}-2]$.

\quad $(ii)$ If $m+|B|=n\geq 5$ and $\text{max}{m,|B|} \leq n-2$, then the code $\mathcal{C}_f$ is minimal violating the condition of Aschikhmin-Barg.
}
\end{theorem}

\begin{theorem}{\rm\cite[Theorems 5.7 and 6.4]{HKWY}\label{HKWYthm5.7}
Let $\Bbb H(m,n)$ be a hierarchical poset with two levels. Let $I_1=A_1 \cup B_1$ and $I_2=A_2\cup B_2$ be two distinct order ideals of $\Bbb H(m,n)$, where $A_i\subseteq [m]$, $B_i\subseteq [n]\backslash[m]$, $i=1,2$ and $I_1 \not\subseteq I_2$, $I_2 \not\subseteq I_1$. Set $\mathcal{I}={I_1,I_2}$.

$(1)$ If $B_1=B_2=\emptyset$, then the weight distribution of the code $\mathcal{C}_f$ is given in \cite[Table 7]{HKWY}.

$(2)$ If $B_1\neq\emptyset$ and $B_2\neq\emptyset$, then the weight distribution of the code $\mathcal{C}_f$ is given in \cite[Table 8]{HKWY}. Moreover, if $B_1\cap B_2=\emptyset$, $|B_1|+|B_2|=n-m$ and $\text{max}\{|B_1|,|B_2|\}\leq n-2$, then the code $\mathcal{C}_f$ is minimal violating the condition of Aschikhmin-Barg with parameters $[2^n-1,n+1,2^{m}-3+2^{|B_1|}+2^{|B_2|}]$.
}
\end{theorem}

\subsection{Binary codes from set operations on simplicial complexes}




%
Let $T=\sum_{j=\alpha}^{\beta}b_j2^{j}$ be a positive integer, where $b_j \in {0,1}$, $b_{\alpha}=b_{\beta}=1$, and $0 \leq \alpha < \beta$. Define
$$l(T)=\sum_{j=\alpha}^{\beta}b_j,~ v_2(T)=\alpha,$$
where $v_2$ denotes the $2$-adic valuation of $T$.

\begin{lemma}{\rm \cite[Lemma 18]{HLL}~Let $T$ and $n$ be positive integers. Then
$$\sum_{i=0}^{n-1}=\bigg\lceil \frac{2^{n-1}+1-T}{2^i} \bigg\rceil = 2^n+l(T)-2T+v_2(T)$$
and
$$\sum_{i=0}^{n-1}=\bigg\lceil \frac{2^{n-1}-T}{2^i} \bigg\rceil = 2^n+l(T)-2T-1.$$

}
\end{lemma}

The defining set $D$ is a subset of $\mathbb{F}_2^{n^*}$, and the binary code $\mathcal{C}_D$ in Equation \eqref{eq2.1}, associated with $D$, is studied. Hyun {\em et al.}  \cite{HLL}  obtained infinite families of binary linear codes $\mathcal{C}_{\Delta^c}$ associated with a simplicial complex $\Delta$, generated by a set of maximal elements $\mathcal{F}$, which includes more than one maximal element. They determined the weight distributions of the binary code $\mathcal{C}_{\Delta^c}$ when $\mathcal{F}$ consists of exactly two elements.

Some results on optimal codes are documented.

\begin{theorem}{\rm\cite[Theorem 20]{HLL}\label{HLLthm20}
~Let $n$ and $s \geq 2$ be integers. Let $\Delta$ be a simplicial complex of $\Bbb F_2^n$ with the set $\mathcal{F} = {A_1, A_2, \ldots, A_s}$ of maximal elements of $\Delta$ and $T=\sum_{i=1}^{s}2^{|A_i|-1}$. Assume that $A_i \backslash \cup_{j \in[s]\backslash{i}}A_j \neq \emptyset$ for any $i \in [s]$ and $\sum_{i=1}^s2^{A_i} < 2^n$. 

$(1)$ The $C_{\Delta^c}$ is a $[2^n-|\Delta|,n,2^{n-1}-T]$ code. If $\cup_{i=1}^sA_i=[n]$ then the minimum distance of its dual is
\begin{equation*}
d((C_{\Delta^c})^{\bot})=
\begin{cases}
4 &\mbox{if}~ s=2,~ |A_1\backslash A_2|=1 ~\mbox{ and }~ |A_2\backslash A_1|\geq 2,\\
3 &\mbox{otherwise}.
\end{cases}
\end{equation*}

$(2)$ The linear code $C_{\Delta^c}$ is length-optimal provided that $|\Delta| \geq 2T-l(T)-v_2(T)$.

$(3)$ $C_{\Delta^c}$ is a Griesmer code if and only if $\mathcal{F}$ is a partition of $\cup_{i=1}^sA_i$ such that the sizes of maximal elements of $\Delta$ are all different.

}
\end{theorem}

\begin{theorem}{\rm\cite[Theorem 27]{HLL}\label{HLLthm27}
~Let $n \geq 2$ be an integer and $\Delta$ a simplicial complex of $\Bbb F_2^n$ with maximal elements ${A_1,A_2}$, such that $2^{|A_1|}+2^{|A_2|}<2^n$. 

$(1)$ If $A_1\cap A_2 =\emptyset$ and $|A_1| < |A_2|$, then $C_{\Delta^c}$ is a length-optimal code with parameters $[2^n-2^{|A_1|}-2^{|A_2|}+1,n,2^{n-1}-2^{|A_1|-1}2^{|A_2|-1}]$. 

$(2)$ If $A_1\cap A_2 =\emptyset$ and $|A_1| = |A_2|$, then $C_{\Delta^c}$ is an optimal code with parameters $[2^n-2^{|A_1|}+1,n,2^{n-1}-2^{|A_1|}]$. 

$(3)$ If $A_1\cap A_2 \neq\emptyset$ and $|A_1| < |A_2|$, then $C_{\Delta^c}$ is a code with parameters $[2^n-2^{|A_1|}-2^{|A_2|}+2^{|A_1\cap A_2|},n,2^{n-1}-2^{|A_1|-1}-2^{|A_2|-1}]$. 

$(4)$ If $A_1\cap A_2 \neq\emptyset$ and $|A_1| = |A_2|$, then $C_{\Delta^c}$ is a code with parameters $[2^n-2^{|A_1|}+2^{|A_1\cap A_2|},n,2^{n-1}-2^{|A_1|}]$. It is optimal when $|A_1| \geq 2^{|A_1\cap A_2|}$.

}
\end{theorem}

\begin{theorem}{\rm\cite[Theorem 5]{WL}\label{WLthm5}
Let $m \geq 3$ be a positive integer, $A,B$ be two elements of $\mathbb{F}_2^m$ such that $B \subset A$, and $D=\Delta_A \setminus \Delta_B.$ 

$(1)$ If $|B|=0$, then the code $\mathcal{C}_D$ is a $[2^{|A|}-1,|A|,2^{|A|-1}]$ one-weight code meeting the Griesmer bound. Its weight distribution is given by $1+{2^{|A|-1}}z^{2^{|A|-1}}$.

$(2)$ If $|B| \geq 0$, the code $\mathcal{C}_D$ is a $[2^{|A|}-2^{|B|}, |A|, 2^{|A|-1}-2^{|B|-1}]$ two-weight code meeting the Griesmer bound. Its weight enumerator is expressed as $1+(2^{|A|-|B|}-1)z^{2^{|A|-1}}+{(2^{|A|}-2^{|A|-|B|})}z^{2^{|A|-1}-2^{|B|-1}}$.

}
\end{theorem}

\begin{theorem}{\rm\cite[Theorem 8]{WL}\label{WLthm8}
Let $m \geq 3$ be a positive integer. Let $A,B$ be two distinct elements of $\mathbb{F}_2^m$ such that $0\leq |B|\leq |A|$ and $A \cap B=\emptyset$. Let $D=(\Delta_A \cup \Delta_B)\setminus\{0\}$. Then $\mathcal{C}_D$ is a $[2^{|A|}+2^{|B|}-2,|A|+|B|,2^{|B|-1}]$ three-weight code with weight enumerator
$$1+(2^{|B|-1})z^{2^{|B|-1}}+(2^{|A|-1})z^{2^{|A|-1}}+(2^{|B|-1})(2^{|A|-1})z^{2^{|A|-1}+2^{|B|-1}}.$$

}
\end{theorem}
%



\begin{theorem}{\rm\cite[Theorems 4-7]{KZL}\label{KZLthm4567}
Let $t_1$ and $t_2$ be positive integers. Define $D_{1}=\Delta_{A}\setminus\Delta_{B}\subseteq\mathbb{F}_{2}^{t_{1}}$, where $\Delta_{A}$, $\Delta_{B}$ are two distinct simplicial complexes of $\mathbb{F}_{2}^{t_{1}}$ with $B\subset A$ and $D_{2}=\{\mathcal{R}_{1},\mathcal{R}_{2},\ldots,\mathcal{R}_{t_{2}}\}$. Then the defining set $D=\{(\mathbf{d}_{1},\mathbf{d}_{2}):\mathbf{d}_{1}\in D_{1},\mathbf{d}_{2}\in D_{2}\}\subseteq{\mathbb{F}}_{2}^{t_{1}+t_{2}}$, and its complement $D^{c*}=(\mathbb{F}_{2}^{t_{1}+t_{2}}\setminus\{\mathbf{0}\})\setminus{D}$. Under this defined set construction, the author presents four classes of infinite families of binary self-orthogonal codes, each with a determined weight distribution, and obtains a class of optimal binary codes.

}
\end{theorem}



\begin{theorem}{\rm \cite[Theorems 3,7]{KZL}\label{KZLthm7}
Suppose that $A$ and $B$ are two elements of $\mathbb{F}_{2}^{t_{1}}$ with $B\subset A$. Let $D_{1}=\Delta_{A}\backslash\Delta_{B}\subseteq \mathbb{F}_{2}^{t_{1}}$, $D_{2}=\{1\}\subseteq\mathbb{F}_{2}$, $D=\{(\mathbf{d}_1,\mathbf{d}_2):\mathbf{d}_1\in D_1,\mathbf{d}_2\in D_2\}$ and $D^{c*}=(\mathbb{F}_{2}^{t_{1}+t_{2}}\backslash\{{\bf 0}\})\backslash D$. The dual code of the code $\mathcal{C}_D$ has minimum distance 4, if $|A|$ and $|B|$ satisfy any one of the following three conditions:\\
$(1)$ $|A| = |B|+1\geq3$,\\
$(2)$ $|A| = |B|+2\geq3$,\\
$(3)$ $|A| \geq |B|+3$.

Moreover, if $|A| = |B|+1\geq3$ or $|A|\geq\max\{|B|+2,3\}$, the code $\mathcal{C}_{D}^\perp $ is an optimal binary code with respect to the Sphere Packing Bound.

}
\end{theorem}

\subsection{Binary codes from posets}

In \cite{HKWY}, the authors apply simplicial complexes to arbitrary posets over $\mathbb{F}_2$. Particularly, they establish a correspondence between anti-chains and simplicial complexes, and determine the weight distributions of binary linear codes associated with hierarchical posets  consisting of two levels.

Let $\Bbb P$ be a poset on $[n]$ and $D=(\mathcal{I}(\Bbb P)^c)$ considered as the complement of $\mathcal{I}(\Bbb P)$ in $2^{[n]}$, where  $\mathcal{I}=\{I_1,\cdots, I_k\}\subseteq \mathcal{O}_{\Bbb P}$. Then the code $\mathcal{C}_D$ in \eqref{eq2.1} with different  choices $D$ is discussed in \cite{HKWY}.

\begin{theorem}{\rm\cite[Theorems 5.1 and 6.1]{HKWY}\label{HKWYthm5.1}
Let $\Bbb H(m,n)$ be a hierarchical poset with two levels. Let $I=A \cup B$ an  order ideal of $\Bbb H(m,n)$, where  $A\subseteq [m]$, $B\subseteq [n]\backslash[m]$. Set $\mathcal{I}=\{I\} $.

$(1)$ If $B=\emptyset$, then the length of the code $\mathcal{C}_D$ is $2^n-2^{|A|}$ and its weight distribution is given in \cite[Table 1]{HKWY}.

$(2)$ If $B\neq\emptyset$, then the length of the code $\mathcal{C}_D$ is $2^n-2^m-2^{|B|}+1$ and its weight distribution is given in \cite[Table 2]{HKWY}. 

In the subcase $(2)$, it is important to note that:

\quad $(i)$ If $B=1<m-1\leq n-2$, then the code $\mathcal{C}_D$ is a Griesmer code with parameters $[2^n-1-2^m,n,2^{n-1}-1-2^{m-1}]$.

\quad $(ii)$ If $m=1$ and $B=n-1$, then the code $\mathcal{C}_D$ is a Griesmer code with parameters $[2^{n-1}-1,n-1,2^{n-2}]$.
 
\quad $(iii)$ The code $\mathcal{C}_D$ is minimal if and only if $(m,|B|)\not\in\{ (1,n-1),(1,n-2),(2,n-2),(n-1,1)\}$ with respect to the condition of Aschikhmin-Barg.
}
\end{theorem}

\begin{theorem}{\rm\cite[Theorems 5.3 and 6.4]{HKWY}\label{HKWYthm5.3}
Let $\Bbb H(m,n)$ be a hierarchical poset with two levels. Let $I_1=A_1 \cup B_1$ and $I_2=A_2\cup B_2$ be two distinct order ideals of $\Bbb H(m,n)$, where  $A_i\subseteq [m]$, $B_i\subseteq [n]\backslash[m]$, $i=1,2$ and $I_1 \not\subseteq I_2$, $I_2 \not\subseteq I_1$.  Set $\mathcal{I}=\{I_1,I_2\} $.

$(1)$ If $B_1=B_2=\emptyset$, then the length of the code $\mathcal{C}_D$ is $2^n-2^{|A_1|}-2^{|A_2|}-2^{|A_1\cap A_2|}$ and its weight distribution is given in \cite[Table 3]{HKWY}.

$(2)$ If $B_1\neq\emptyset$ and $B_2\neq\emptyset$, then the length of the code $\mathcal{C}_D$ is $2^n-2^{|B_1|}-2^{|B_2|}-2^{|B_1\cap B_2|}$ and its weight distribution is given in \cite[Table 4]{HKWY}. 

For the subcase $(2)$, if $|B_1|=|B_2|=1$ and $1 < m \leq n-2$, then the code $\mathcal{C}_D$ is a distance-optimal linear code with parameters $[2^n-2^m-2,n,2^{n-1}-2^{m-1}-2]$. It is minimal when $n \geq 4$ with respect to the condition of Aschikhmin-Barg.
}
\end{theorem}

In \cite{WL2}, the authors utilized hierarchical posets consisting of two levels to construct binary linear codes. They investigate the necessary and sufficient conditions for these codes to be self-orthogonal. Furthermore, they leverage these self-orthogonal codes to derive many interesting binary quantum codes, which include many binary quantum Hamming codes.

\begin{theorem}{\rm\cite[Theorem 1 and Corollary 1]{WL2}\label{WL2thm1}
Let $\Bbb H(m,n)$ be a hierarchical poset with two levels. Let $I_1=A_1 \cup B_1$ and $I_2=A_2\cup B_2$ be two distinct order ideals of $\Bbb H(m,n)$, where  $A_i\subseteq [m]$, $B_i\subseteq [n]\backslash[m]$, $i=1,2$ and $I_2 \subset I_1$. Let $D = (I_1(\Bbb P)\backslash(T_2(\Bbb P)))$.

$(1)$ If $B_1 = \emptyset$, then the code $\mathcal{C}_D$ has parements$[2^{|A_1|}-2^{|A_2|}, |A_1|]$ and its weight distribution is given in \cite[Table 1]{WL2} .

$(2a)$ If $B_1 \neq \emptyset$, $B_2 = \emptyset$ and $|A_2|=m$, then the code $\mathcal{C}_D$ has parements$[2^{|B_1|}-1, 1+|B_1|, 2^{|A_1|-1}-1]$ and its weight distribution is given in \cite[Table 2]{WL2}.

$(2b)$ If $B_1 \neq \emptyset$, $B_2 = \emptyset$ and $|A_2|<m$, then the code $\mathcal{C}_D$ has parements$[2^m+2^{|B_1|}-2^{|A_2|}-1, m+|B_1|]$ and its weight distribution is given in \cite[Table 3]{WL2}.

$(3)$ If $B_1 \neq \emptyset$, $B_2 \neq \emptyset$, then the code $\mathcal{C}_D$ has length $2^{|B_1|}-2^{|B_2|}$ and its weight distribution is given in \cite[Table 4]{WL2}.

For the subcase $(2a)$, note that the code $\mathcal{C}_D$ has parameters $[2^{|B_1|}-1, 1+|B_1|, 2^{|A_1|-1}-1]$ and meets the Griesmer bound with equality.

For the subcase $(3)$, note that:

\quad $(i)$ If $|B_1| = |B_2| +1$, then the code $\mathcal{C}_D$ has parameters $[2^{|B_2|},2+|B_2|,2^{|B_2|}]$ and meets the Griesmer bound with equality.

\quad $(ii)$ If $|B_2| = 1$ and $3\leq |B_1|$, then the code $\mathcal{C}_D$ has parameters $[2^{|B_1|}-2,1+|B_1|,2^{|B_1|-1}-2]$ and it is almost optimal.

}
\end{theorem}

In \cite{WHY2}, authors made an extension to the method of using simplicial complexes by applying it to posets over $\Bbb F_2$.

\begin{theorem}{\rm\cite[Theorems 3.1, 4.1 and 4.2]{WHY2}\label{WHYthm3.1}
~Let $I$ be a down set of $\Bbb P=(m \oplus n, \preccurlyeq)$, and $D=\Bbb F_2^n\backslash I(\Bbb P)$.

$(1)$ When $I=[i]$ and $1\leq i \leq m$, then the weight distribution of $\mathcal{C}_D$ is presented in \cite[Table \uppercase\expandafter{\romannumeral 1}]{WHY2}  with parameters $[2^n-i-1,n]$.

For the subcase $(1)$, note that

\quad $(i)$ If $n \geq 2$ is an integer and $i=1$, then the code $\mathcal{C}_D$ with parameters $[2^n-2,n,2^{n-1}-2]$ is a
Griesmer code.

\quad $(ii)$ If $n>m \geq 2$ is an integer and $i=2$, then the code $\mathcal{C}_D$ with parameters $[2^n-3,n,2^{n-1}-2]$ is optimal.

\quad $(iii)$ If $n \geq 3$ is an integer and $i=3$, then the code $\mathcal{C}_D$ with parameters $[2^n-4,n,2^{n-1}-3]$ is almost optimal.

$(2)$ When $I=[j]\backslash [m]$ and $m+1\leq j \leq n$, then the weight distribution of $\mathcal{C}_D$ is presented in \cite[ Table \uppercase\expandafter{\romannumeral 2}]{WHY2} with parameters $[2^n+m-j-1,n]$.

$(3)$ When $I=[i]\cup ([j]\backslash [m])$ and $1 \leq i \leq m$, then the weight distribution of $\mathcal{C}_D$ is presented in \cite[Table \uppercase\expandafter{\romannumeral 3}]{WHY2}  with parameters $[2^n+m-i-j-1-i(j-m),n]$.

For the subcase $(3)$, note that

\quad $(i)$ If $n \geq 3$ is an integer $m=2$, $i=1$ and  $j=3$, then the code $\mathcal{C}_D$ with parameters $[2^n-4,n,2^{n-1}-2]$ is a Griesmer code.

\quad $(ii)$ If $n \geq 4$ is an integer $m=2$, $i=2$ and  $j=3$, then the code $\mathcal{C}_D$ with parameters $[2^n-6,n,2^{n-1}-4]$ is optimal.

\quad $(iii)$ If $n \geq 4$ is an integer $m\geq i=2$ and $j-m = 2$, then the code $\mathcal{C}_D$ with parameters $[2^n-9,n,2^{n-1}-6]$ is almost optimal.

}
\end{theorem}

\subsection{New Construction Methods Using Simplicial Complexes}

\begin{theorem}{\rm\cite[Theorem 1]{H23}\label{H23thm1}
Let $\Delta$ be a simplicial complex of $\Bbb F_q^m$ with the support $\mathcal{A}=\{A_{1},A_{2},\ldots,A_{h}\}$, where $1\leq|A_{1}|\leq|A_{2}|\leq\cdots\leq |A_{h}|<m$. Assume that $A_i\setminus(\cup_{1\leq j\leq h,j\neq i}A_j)\neq\emptyset $ for any $1\leq i\leq h$ and $q^{m}>\sum_{1\leq i\leq h}q^{|A_{i}|}$. Denote $T=\sum_{1\leq i\leq h}q^{|A_{i}|-1}$. Let $C_{\overline{\Delta}^{c}}$ be defined as in \eqref{eq2.1}. Then the $C_{\overline{\Delta}^{c}}$ has parameters $[(q^m-|\Delta|)/(q-1),m,q^{m-1}-T]$, where $|\Delta|=\sum_{\emptyset\neq S\subseteq\mathcal{A}}(-1)^{|S|-1}q^{|\cap S|}$ and $\cap S$ is defined as $\cap S=\cap_{A\in S}A$.


$(1)$ \cite[Corollary 1]{H23} Assume  $|A_{i}\cap A_{j}|=0$ for any $1\leq i\le j \leq h$, then $C_{\overline{\Delta}^{c}}$  is an at most $2^h$-weight $[(q^{m}-\sum_{i=1}^{h}q^{|A_{i}|}+h-1)/(q-1),m,q^{m-1}-T]$ linear code, 

\quad $(i)$ $C_{\overline{\Delta}^{c}}$ is a Griesmer code if and only if $|A_{i}\cap A_{j}|=0$ for any $1\leq i\le j \leq h$ and at most $q-1$ of $|A_{i}|$’s are the same.

\quad $(ii)$ $C_{\overline{\Delta}^{c}}$ is a near Griesmer code if and only if $\ell(T)=h-(q-1)$.

\quad $(iii)$ $C_{\overline{\Delta}^{c}}$ is distance-optimal if $\ell(T)+(q-1)(\nu(T)+1)>h$. Moreover, when $|A_{i}|={\boldsymbol{\varepsilon}}$ for $1\leq i\leq h$, where $\varepsilon$ is a positive integer, it is distance-optimal if  $\ell(h)+(q-1)(\nu(h)+\mathbf{\varepsilon})>h$.

$(2)$ $C_{\overline{\Delta}^{c}}$ is distance-optimal if $|\Delta|-1+(q-1)(\nu(T)+1)> qT-\ell(T)$.

All the codes' weight distributions were given in \cite{H23}. The Griesmer codes in  Theorem \ref{H23thm1} are indeed the Solomon-Stiffler codes\cite[Remark 5]{H23}.
}
\end{theorem}

\begin{theorem}{\rm\cite[Theorem 2]{H23}\label{H23thm2}
Let $\Delta$ be a simplicial complex of $\Bbb F_q^m$ with the exactly one maximal element and its support is $\{A\}$ with $A \subseteq [m]$ and $1\leq|A|<m$. Then $C_{\overline{\Delta}^{c}}$ defined by \eqref{eq2.1} is a $2$-weight $[(q^{m}-q^{|A|})/(q-1),m,q^{m-1}-q^{|A|-1}]$ linear code and it is a Griesmer code. Also it is a Solomon-Stiffler code \cite[Remark 6]{H23}.

}
\end{theorem}

\begin{theorem}{\rm\cite[Theorem 3]{H23}\label{H23thm3}
Let $\Delta$ be a simplicial complex of $\Bbb F_q^m$ with the  support $\mathcal{A}=\{A_{1},A_{2}\}$, where $1\leq|A_{1}|\leq|A_{2}|<m$. Assume that $q^{m}>q^{|A_{1}|}+q^{|A_{2}|}$. Let $T=q^{|A_{1}|-1}+q^{|A_{2}|-1}$. Then $C_{\overline{\Delta}^{c}}$ defined by \eqref{eq2.1} is an at most $5$-weight $[(q^{m}-q^{|A_{1}|}-q^{|A_{2}|}+q^{|A_{1}\cap A_{2}|})/(q-1),m,q^{m-1}-q^{|A_{1}|-1}-q^{|A_{2}|-1}]$   linear code. Moreover,

$(1)$ When $|A_1 \cap A_2| = 0$  and $|A_1| = |A_2|$, $C_{\overline{\Delta}^{c}}$
is a near Griesmer code (also distance-optimal) if $q=2$ and it is a Griesmer code if $q>2$. It reduces to a $3$-weight code in this case.

$(2)$ When $|A_1 \cap A_2| = 0$ and $|A_1| < |A_2|$, $C_{\overline{\Delta}^{c}}$ is a Griesmer code and it reduces to a $4$-weight code.

$(3)$ When $|A_1 \cap A_2| > 0$ and $|A_1| = |A_2|$, $C_{\overline{\Delta}^{c}}$ is distance-optimal if $\ell(T)+(q-1)(\nu(T)+1)>q^{|A_{1}\cap A_{2}|}+1$ and it reduces to a $4$-weight code. Specially, $C_{\overline{\Delta}^{c}}$ is a near Griesmer code if $q>2$ and $|A_1 \cap A_2| = 1$.

$(4)$ When $|A_1 \cap A_2| > 0$ and $|A_1| < |A_2|$, $C_{\overline{\Delta}^{c}}$ is distance-optimal if $ (q-1)|A_{1}|+1>q^{|A_{1}\cap A_{2}|} $. Specially, $C_{\overline{\Delta}^{c}}$ is a near Griesmer code if $|A_1 \cap A_2| = 1$.

}
\end{theorem}

\begin{theorem}{\rm\cite[Theorem 4]{H23}\label{H23thm4}
Let $\Delta$ be a simplicial complex of $\Bbb F_q^m$ with the  support $\mathcal{A}=\{A_{1},A_{2},A_{3}\}$, where $1\leq|A_{1}|\leq|A_{2}|\leq|A_{3}|<m$. 

Assume that $A_i\setminus(\cup_{1\leq j\leq3,j\neq i}A_j)\neq\emptyset  $ for any $1\leq i\leq 3$ and $q^{m}>\sum_{1\leq i\leq3}q^{|A_{i}|}$. Let $T=\sum_{1\leq i\leq3}q^{|A_{i}|-1}$
. Then $C_{\overline{\Delta}^{c}}$ defined by \eqref{eq2.1} is an at most $19$-weight $[(q^{m}-|\Delta|)/(q-1),m,q^{m-1}-T]$ linear code

$(1)$ $C_{\overline{\Delta}^{c}}$ is a Griesmer code if and only if $|A_i \cap A_j| = 0$ for
$1 \leq i < j \leq 3$ and at most $q-1$ of $|A_i|$’s are the same (which always holds for $q > 3$).

$(2)$ $C_{\overline{\Delta}^{c}}$ is a near Griesmer code if one of the followings
holds:
 
\quad $(i)$ $|A_i \cap A_j| = 1$ for only one element $(i,j)$ in the set $\{(i,j) : 1 \leq i < j \leq 3 \}$ and $|A_i \cap A_j| = 0$ for the other two $(i,j)$’s, and at most $q-1$ of $|A_i|$’s are the same,

\quad $(ii)$ $q=3$ , $|A_i \cap A_j| = 0$ for $1 \leq i < j \leq 3$, and $|A_{1}|=|A_{2}|=|A_{3}|$, and

\quad $(iii)$ $q=2$, $|A_i \cap A_j| = 0$ for $1 \leq i < j \leq 3$, and $|A_{1}|=|A_{2}|<|A_{3}|-1$ or $|A_{1}|\leq|A_{2}|=|A_{3}|$.

$(3)$ $C_{\overline{\Delta}^{c}}$ is distance-optimal if $(q-1)(\nu(T)+1)+\ell(T)-1> \sum_{1\leq i<j\leq3}q^{|A_{i}\cap A_{j}|}-q^{|A_{1}\cap A_{2}\cap A_{3}|}$.


}
\end{theorem}

\subsection{Codes over $\Bbb F_4 $}

As we know, $\Bbb F_4\cong \Bbb F_2[x]/\langle x^2+x+1\rangle$, where $x^2+x+1$ is the only irreducible polynomial of degree two in $\Bbb F_2[x]$. Let $w$ be an element in some extended field of $\Bbb F_2$ such that $w^2+w+1=0$. Then $\Bbb F_4=\Bbb F_2(w)$ and for each $u \in \mathbb F_4$ there is a unique representation $u = a + wb$, where $a, b \in \Bbb F_2$. Let $m$ be a positive integer, and $\mathbb F_4^m$ be the set of $m$-tuples over $\mathbb F_4$. Any vector $\mathbf x \in \mathbb F_4^m$ can be written as $\mathbf x =\mathbf a + w\mathbf b$, where $\mathbf a,\mathbf b\in\mathbb F_2^m$.

Let the defining set $D = \Delta_1 +w\Delta_2 \subseteq \Bbb F_4^m$ be the defining set of the code $\mathcal C_D$ in \eqref{eq2.1}. Then the quaternary codes were investigated in \cite{ZW} and \cite{WLX}.

\begin{theorem}{\rm\cite[Theorem 3.1]{ZW}\label{ZWthm3.1}
~Let $A,B$ be two subsets of $[m]$ and $D = \Delta_A^c + w\Delta_B \subset \Bbb F_4^m$. Then $\mathcal C_D$ is a $[(2^m-2^{|A|})2^{|B|},m]$ quaternary code and its weight distribution is presented in \cite[Table 1]{ZW}.

}
\end{theorem}

\begin{corollary}{\rm\cite[Corollary 3.3]{ZW}
~Let $A$ be a subset of $[m]$ and $D = \Delta_A^c + w\Bbb F_2^m \subset \Bbb F_4^m$. Then $\mathcal C_D$ is a $[(2^m-2^{|A|})2^{|B|},m, 3\times2^{2^m-2}-3\times2^{|A|+m-2}]$ two-weight quaternary Griesmer code and it is minimal when $A$ is a proper subset of $[m]$ with respect to the condition of Ashikhmin-Barg. 

}
\end{corollary}

%


%
 Wu \emph{et al.} \cite{WLX} studied the connection between quaternary codes and their binary subfield codes. The authors constructed the quaternary code $\mathcal{C}_{D}$ for simplicial complexes generated by both a single maximal element and two maximal elements.

\begin{theorem}{\rm \cite[Proposition 4.2 and Theorem 4.4]{WLX}\label{WLXpro4.2}
~Let $A,B$ be two subsets of $[m]$ and $D=\Delta_A+w\Delta_B\subset \Bbb F_4^m$. 

$(1)$ Then $\mathcal{C}_{D^{*}}$ is a $[2^{|A|+|B|}-1, |A\cup B|, 2^{|A|+|B|-1}]$ quaternary code and its weight distribution is presented in \cite[Table 2]{WLX}.

$(2)$ Then $\mathcal{C}_{D^{c}}$ is a $[4^{m}-2^{|A|+|B|}, m,3\times 2^{2m-2}-3\times2^{|A|+|B|-2}]$ quaternary code and its weight distribution is presented in \cite[Table 6]{WLX}. Moreover, the code $\mathcal{C}_{D^{c}}$ is a Griesmer code.

}
\end{theorem}

\subsection{ Codes over $\Bbb F_{3}$}

%

When $\Delta_1$ and $\Delta_2$ are two disjoint simplicial complexes of $\mathbb{P}([n])$,  where $\mathbb{P}([n])$ denotes the power set of $[n]$, consider the set 
$$\mathcal{D}_2(\Delta_1,\Delta_2)=\{(A,B):A\in\Delta_1,B\in\Delta_2\}.$$
Since $\Delta_1\cap\Delta_2=\emptyset$, we have $\mathcal{D}_2(\Delta_1,\Delta_2)\subseteq\mathcal{D}_2([n])$. Considering the vector space $\mathbb{F}_3^n$,
there exists a bijection
$$\begin{array}{rcl}\varphi=(\varphi_1,\varphi_2):\mathbb{F}_3^n&\longrightarrow&\mathcal{D}_2([n])\\u=(u_1,u_2,\ldots,u_n)&\mapsto&(\varphi_1(u),\varphi_2(u))\end{array}
$$
where $\varphi_{1}(u)=\{i:u_{i}=1\}$ and $\varphi_{2}(u)=\{j:u_{j}=-1\}$. The set $\mathcal{D}_2(\Delta_1,\Delta_2)$ given by two disjoint simplicial complexes, under the map $\varphi$, will then be identified with the subset of $\mathbb{F}_3^n$ without any real ambiguity \cite{PL}.

\begin{theorem}{\rm\cite[Theorem 4.3]{PL}
~Let $\Delta_1=\langle\{r\},\{s\}\rangle$ and $\Delta_2=\langle\{t\}\rangle$ be simplicial complexes of $(\Bbb P[n])$, where $1\leq r,s,t \leq n$ are pairwise distinct and $n \geq 3$. Then $\mathcal{C}_{\mathcal{D}_2(\Delta_1,\Delta_2)^c}$ is an optimal $[3^n-8,n,3^n-3^{n-1}-6]$ code, and its weight distribution is given in \cite[Table 3]{PL}.
}
\end{theorem}

\subsection{Codes over $\Bbb F_{2^n}$ with $n\ge 3$}


%
Let $\Bbb F_8 = \Bbb F_2(w)$, where $w\in \Bbb F_8$ satisfies the polynomial $y^3+y+1=0$. Let $m$ be a positive integer. For each $\mathbf v \in \mathbb F_8^m$, there exists a unique $\mathbf a,\mathbf b,\mathbf c\in\mathbb F_2^m $ such that $\mathbf v =\mathbf a + w\mathbf b +w^2\mathbf c$. This structure of $\Bbb F_8$ (or $\Bbb F_{2^3}$) is used in \cite{SS,SS5}, and in \cite{LY}, authors extended this to $\Bbb F_{2^n}$.

\begin{theorem}{\rm\cite[Theorem 3.6]{SS5}\label{SS5thm3.6}
~Let $L,M,N$ be subsets of $[m]$ such that $L \cup M\cup N \subsetneq [m]$. Further, assume that at least two of the sets $L \backslash (M\cup N)$, $N \backslash (M\cup L)$, and $M \backslash (L\cup N)$ are nonempty. Let $D=\Delta_L+w\Delta_M+w^2\Delta_N$. Then $\mathcal{C}_{D^c}$ has parameters $[2^{3m}-2^{|L|+|M|+|N|},m,7\times2^{3(m-1)}-7\times2^{|L|+|M|+|N|-3}]$ and it has codewords of weights 0, $7\times2^{3(m-1)}-7\times2^{|L|+|M|+|N|-3}$, $7\times2^{3(m-1)}-6\times2^{|L|+|M|+|N|-3}$, $7\times2^{3(m-1)}-4\times2^{|L|+|M|+|N|-3}$, and $7\times2^{3(m-1)}$. Its weight distribution can be obtained in \cite{SS5}. The $\mathcal{C}_{D^c}$ is a Griesmer code and hence it is distance optimal. Furthermore, it is a minimal code if $|L|+|M|+|N| \leq 3m-4$.
}
\end{theorem}

\begin{theorem}{\rm\cite[Theorem 3.7]{SS5}\label{SS5thm3.7}
~Let $\emptyset\neq L=M=N\subsetneq [m]$ and $D=\Delta_L+w\Delta_L+w^2\Delta_L$. Then $\mathcal{C}_{D^c}$ has parameters $[2^{3m}-2^{3|L|},m,7\times2^{3(m-1)}-7\times2^{3|L|-1}]$ and it has codewords of weights 0, $7\times2^{3(m-1)}-7\times2^{3|L|-1}$, and $7\times2^{3(m-1)}$. The $\mathcal{C}_{D^c}$ is a Griesmer code and hence it is distance optimal. Further, it is a minimal code if $3(m-|L|)-4\geq 0$.
}
\end{theorem}

%


%
In \cite[Proposition 3.3]{SS}, the authors established the connection between the weight distribution of the constructed codes in $\Bbb F_{2^i}$ (where $i=1,2,3,4$) and the frequency of codewords weights. Building upon this observation, the authors formulated some conjectures to extend this relationship to $\Bbb F_{2^n}$. In \cite{LY}, Liu and Yu provided a comprehensive proof of these conjectures by leveraging the relevant conclusions of linear feedback shift register (LFSR).


\begin{theorem}{\rm\cite[Theorem 3.4]{SS}\label{SSthm3.4}
~Let $\emptyset\neq L=M=N\subsetneq [m]$ and $D=\Delta_L+w\Delta_L+w^2\Delta_L$. Then $\mathcal{C}_{D^*}$ is a $1$-weight octanary linear code over $\Bbb F_8$ of length $2^{3|L|}-1$, dimension $|L|$, and distance $7 \cdot 2^{3(|L|-1)}$. $\mathcal{C}_{D^*}$ is a Griesmer code and hence it is a distance optimal. Its weight distribution is given in \cite[Theorem 3.4]{SS}.

}
\end{theorem}

\begin{theorem}{\rm\cite[Theorem 3.9]{SS}\label{SSthm3.9}
~Let $L,M,N$ be subsets of $[m]$ such that $L \cup M\cup N \subsetneq [m]$. Further, assume that at least two of the sets $L \backslash (M\cup N)$, $N \backslash (M\cup L)$, and $M \backslash (L\cup N)$ are nonempty. Let $D=\Delta_L+w\Delta_M+w^2\Delta_N$. Then $\mathcal{C}_{D^c}$ is a $4$-weight octanary linear code over $\Bbb F_8$ of length $2^{3m}-2^{|L|+|M|+|N|}$, dimension $m$, and distance $7 \cdot 2^{3(m-1)}- 7\cdot2^{|L|+|M|+|N|-3}$. In particular, $\mathcal C_{D^c}$ has codewords of weights $0$, $7 \cdot 2^{3(m-1)}- 7\cdot2^{|L|+|M|+|N|-3}$, $7 \cdot 2^{3(m-1)}- 6\cdot2^{|L|+|M|+|N|-3}$, $7 \cdot 2^{3(m-1)}- 4\cdot2^{|L|+|M|+|N|-3}$, and $7 \cdot 2^{3(m-1)}$. $\mathcal C_{D^c}$ is a Griesmer code and hence it is a distance optimal.

}
\end{theorem}

\begin{theorem}{\rm\cite[Theorem 3.10]{SS}\label{SSthm3.10}
~Let $\emptyset\neq L=M=N\subsetneq [m]$ and $D=\Delta_L+w\Delta_L+w^2\Delta_L$. Then $\mathcal{C}_{D^c}$ is a $2$-weight octanary linear code over $\Bbb F_8$ of length $2^{3m}-2^{3|L|}$, dimension $m$, and distance $7\times2^{3(m-1)}-7\times2^{3(|L|-1)}$. In particular, $\mathcal C_{D^c}$ has codewords of weights $0$, $7\times2^{3(m-1)}-7\times2^{3(|L|-1)}$, and $7\times2^{3(m-1)}$. $\mathcal C_{D^c}$ is a Griesmer code and hence it is a distance optimal. Further, it is a minimal code if $3(m-|L|)-4\geq 0$.
}
\end{theorem}

%


%

In response to the questions and conjectures put forth in \cite{SS}, the authors of \cite{LY} undertook a comprehensive investigation into linear codes over $\Bbb F_{2^n}$, along with their corresponding binary subfield codes. They utilized a result derived from LFSR sequences, as presented in \cite[Definition 2.1]{LY}, to establish a meaningful relationship between the weights of codewords in two specific codes that were derived from simplicial complexes \cite[Proposition 3.1]{LY}. By employing this approach, they were able to provide a compelling and comprehensive proof for the conjectures proposed in \cite{SS}.

\begin{theorem}{\rm\cite[Theorem 4.1]{LY}\label{LYthm4.1}
~Let $L$ be a nonempty subset of $[m]$. Consider the defining set $D=\Delta_L+w\Delta_L+\cdots+w^{n-1}\Delta_L \subseteq \Bbb F_{2^n}^m$. Then the code $\mathcal{C}_{D^*}$ is a $1$-weight linear code over $\Bbb F_{2^n}^m$ of length $2^{n|L|}-1$, dimension $|L|$, and distance $(2^n-1) 2^{n(|L|-1)}$. It is a Griesmer code. Moreover, it is a minimal code.

}
\end{theorem}

\begin{theorem}{\rm\cite[Theorem 4.2]{LY}\label{LYthm4.2}
~Let $L_i$ be nonempty subsets of $[m]$ such that at least one subset
is proper. Let $R_i=L_i\backslash \cup_{j\neq i}L_i$ and suppose that $R_0,R_1,\cdots,R_{n-2}$ are nonempty. Let $D = \Delta_{L_0} + w\Delta_{L_1} + \cdots + w^{n-1}\Delta_{L_{n-1}} \subseteq \Bbb F_{2^n}^m$, $\mathcal{C}_{D^c}$ is a $(n+1)$-weight linear code over $\Bbb F_{2^n}$ of length $2^{nm}-2^{\sum_{i=0}^{n-1}|L_i|}$, dimension $m$, and distance $(2^n-1)(2^{n(m-1)}- 2^{\sum_{i=0}^{n-1}|L_i|-n})$. In particular, $\mathcal{C}_{D^c}$ has nonzero codewords of weights $(2^n-1)2^{n(m-1)}- 2^{\sum_{i=0}^{n-1}|L_i|-n}(2^n-2^i)$, $0 \leq i \leq n$. It is a Griesmer code, hence it is distance optimal. In fact, it is a minimal code if $\sum_{i=0}^{n-1}|L_i|\leq nm-(n+1)$.

}
\end{theorem}

\begin{theorem}{\rm\cite[Theorem 4.3]{LY}\label{LYthm4.3}
~Let $L$ be a nonempty subset of $[m]$ and let $D=\Delta_L+w\Delta_L+\cdots+w^{n-1}\Delta_L \subseteq \Bbb F_{2^n}^m$. Then the code $\mathcal{C}_{D^c}$ is a $2$-weight linear code over $\Bbb F_{2^n}$ of length $2^{nm}-2^{n|L|}$, dimension $m$, and distance $(2^n-1)(2^{n(m-1)}- 2^{n(|L|-1)}$. In particular, $\mathcal{C}_{D^c}$ has nonzero codewords of weights $(2^n-1)2^{n(m-1)}- 2^{n(|L|-1)}$ and $(2^n-1)2^{n(m-1)}$. It is a Griesmer code, hence it is distance optimal. In fact, it is a minimal code if $n(m-|L|)\geq n+1$.

}
\end{theorem}

For codes over finite fields, we also have the following problems:

\begin{problem}{\rm Find more optimal codes over finite fields and determine their weight distributions in different metrics by using simplicial complexes or posets. 

}
\end{problem}

\begin{problem}{\rm Find more infinite families of linear minimal  codes over finite fields violating Ashikhmin- Barg’s condition by using simplicial complexes or posets. 
}
\end{problem}

\section{Subfield codes}


In this section, we will consider the subfield codes of some linear codes over finite fields in terms of simplicial complexes.

Let $q=2$, $m=2$, and $\{\alpha_{1}=1,\alpha_{2}=w\}$ be a basis of $\Bbb F_{4}$ over $\Bbb F_{2}$. Assume that $g_{ij}=g_{ij}^{(0)}+wg_{ij}^{(1)}$, where $g_{ij}^{(0)}, g_{ij}^{(1)}\in \Bbb F_{2}$. Hence $\mbox{Tr}_{4/2}( g_{ij} \alpha_{1})=g_{ij}^{(1)}$ and $\mbox{Tr}_{4/2}( g_{ij} \alpha_{2})=g_{ij}^{(0)}+g_{ij}^{(1)}$. The following theorem given in \cite{WLX} examines the relationship between the binary subfield code $C_{D}^{(2)}$ with respect to $C_{D}$.

\begin{theorem}\label{thm3.2} {\rm  \cite[Theorem 3.2]{WLX}  Let $\mathcal C$ be an $[n,k]$ linear code over $\mathbb{F}_{4}$ with a generator matrix $G=G_{1}+wG_{2}$, where $w\in \Bbb F_{4}$ with $w^{2}+w+1=0$ and $G_{1}, G_{2}$ are two matrices over $\Bbb F_{2}$. Then the binary subfield code $\mathcal C^{(2)}$ with respect to $\mathcal C$ has a generator matrix 
$$G^{(2)}=\left(\begin{array}{cclc} G_{2}\\ 
G_{1}+G_{2} \end{array}\right).$$ Moreover, if the quaternary code $\mathcal C$ has the defining set $D=D_{1}+wD_{2}$ with $D_{1}, D_{2} \subseteq \mathbb F_2^m$, then the binary subfield code $\mathcal C^{(2)}$ with respect to $\mathcal C$ has the defining set: 
$$D^{(2)}=\{(\mathbf d_{2}, \mathbf d_{1}+\mathbf d_{2}): \mathbf d_{1}\in D_{1}, \mathbf d_{2}\in D_{2} \}.$$}
\end{theorem}

\begin{theorem}{\rm \cite[Theorem 5.2]{WLX}\label{WLXthm5.2}
~Let $A,B$ be two subsets of $[m]$ and $D=\Delta_A+w\Delta_B\subset \Bbb F_4^m$. Then the subfield code $\mathcal{C}^{(2)}_{D^{c}}$ with respect to $\mathcal{C}_{D^{c}}$ in Theorem \ref{WLXpro4.2}$(2)$ is a $[2^{2m}-2^{|A|+|B|}, 2m,2^{2m-1}-2^{|A|+|B|-1} ]$ two-weight binary linear code and its weight distribution is given by 
$$1+(4^{m}-2^{2m-|A|-|B|})z^{2^{2m-1}-2^{|A|+|B|-1}}+(2^{2m-|A|-|B|}-1)z^{2^{2m-1}}.$$ Moreover, the code $\mathcal{C}^{(2)}_{D^{c}}$ is a Griesmer code.}
\end{theorem}

%


%

%


%
The following theorem, given in \cite{SS} and \cite{SS5}, shows the relationship between the binary subfield code $C_{D}^{(2)}$ and $C_{D}$.

Let $q=2$, $m=3$, and $\{\alpha_{1}=1,\alpha_{2}=w,\alpha_3=w^2\}$ be a basis of $\Bbb F_{8}$ over $\Bbb F_{2}$. Assume that $g_{ij}=g_{ij}^{(1)}+wg_{ij}^{(3)}+w^2g_{ij}^{(3)}$, where $g_{ij}^{(1)}, g_{ij}^{(2)},g_{ij}^{(3)}\in \Bbb F_{2}$. Then $\mbox{Tr}_{8/2}( g_{ij} \alpha_{1})=g_{ij}^{(1)}$, $\mbox{Tr}_{8/2}( g_{ij} \alpha_{2})=g_{ij}^{(3)}$, and $\mbox{Tr}_{4/2}( g_{ij} \alpha_{3})=g_{ij}^{(2)}$.

\begin{theorem}\label{SSthm2.6} {\rm  \cite[Theorem 2.6]{SS} Let $\mathcal B$ be the order basic $\{1<w<w^2\}$ of $\Bbb F_8$ over $\Bbb F_{2}$. Suppose $G:=G_{1}+wG_{2}+w^2G_{3} \in M_{k\times n}(\Bbb F_8)$ generates the linear code $C$, where $G_{i} \in M_{k\times n}(\Bbb F_{2}),1\leq i\leq 3$. Then the binary subfield code $C^{(2)}$ with respect to $C$ has a generator matrix  
$$G^{(2)}=\left(\begin{array}{cclc} G_{1}\\ 
G_{3}\\
G_{2} \end{array}\right).$$

Moreover, if the code $C_{D}$ has the defining set $D=D_{1}+wD_{2}+w^2D_{3}\subseteq \Bbb F_8^m$ with $D_{1}, D_{2},D_{3} \subseteq \mathbb F_2^m$, then the binary subfield code $C_{D}^{(2)}$ with respect to $C_{D}$ has the defining set: 
$$D^{(2)}=\{(\mathbf d_{1}, \mathbf d_{3},\mathbf d_{2}): \mathbf d_{1}\in D_{1}, \mathbf d_{2}\in D_{2}, \mathbf d_{3}\in D_{3} \}.$$
}
\end{theorem}

%


\begin{theorem}{\rm \cite[Theorem 4.1]{SS}\label{SSthm4.1}
~Let $L$, $M$, and $N$ be nonempty subsets of $[m]$. Define $D=\Delta_L+w\Delta_M+w^2\Delta_N$, such that $D^{(2)}=\{(d_1,d_2,d_3):d_1 \in \Delta_L, d_2 \in \Delta_M, d_3 \in \Delta_N  \} \subset (\Bbb F_2^m)^3$. Then the code $\mathcal{C}_{D^*}^{(2)}$ is a $[2^{|L|+|M|+|N|}-1,|L|+|M|+|N|,2^{|L|+|M|+|N|-1}]$ 1-weight binary code, and its weight distribution is given in \cite[Table 1]{SS}. It is an optimal and minimal code.
}
\end{theorem}

\begin{theorem}{\rm\cite[Theorem 4.7]{SS}\label{SSthm4.7}
~Let $L$, $M$, and $N$ be nonempty subsets of $[m]$, with at least one subset being proper. Define $D=\Delta_L+w\Delta_M+w^2\Delta_N$. Then the code $\mathcal{C}_{D^c}^{(2)}$ is a $[2^{3m}-2^{|L|+|M|+|N|},3m,2^{3m-1}-2^{|L|+|M|+|N|-1}]$ 2-weight binary code, and its weight distribution is given in \cite[Table 2]{SS}. It is a Griesmer code, and if $|L|+|M|+|N|\leq 3m-2$, it is also a minimal code.
}
\end{theorem}


\begin{theorem}{\rm\cite[Theorem 4.6]{SS5}\label{SS5thm4.6}
~Let $L$, $M$, and $N$ be subsets of $[m]$, with at least one subset being proper. Define $D=\Delta_L+w\Delta_M+w^2\Delta_N$. Then $\mathcal{C}_{D^c}^{(2)}$ is a two-weight binary code, with its weight distribution given in \cite[Table 2]{SS5}. It is a Griesmer code, hence it is distance optimal. Furthermore, when $|L|+|M|+|N| \leq 3m-2$, it is also a minimal code.
}
\end{theorem}


The following Theorem documented in  \cite{LY} and examined  relationship between the binary subfield code $C_{D}^{(2)}$ with respect to $C_{D}$.

Let $q=2$ and $f(x)=x^n+a_{n-1}x^{n-1}+\cdots+a_1x+a_0 \in \Bbb F_{2}[x]$ be an irreducible polynomial and $w$ a root of $f(x)$. Then $\Bbb F_{2^n}=\Bbb F_{2}(w)$ and $\{ 1,w,\cdots,w^{n-1} \}$ is a basis of $\Bbb F_{2^n}$ over $\Bbb F_{2}$. 
For $G:=G_{0}+wG_{1}+\cdots+w^{n-1}G_{n-1} \in M_{k\times l}(\Bbb F_{2^n})$, where $G_i\in M_{k\times l}(\Bbb F_{2})$. For each entry $g_{ij}$ of $G$, let
$g_{ij}=g_{ij}^{(0)}w+g_{ij}^{(1)}+\cdots+w^{n-1}g_{ij}^{(n-1)}$, where $g_{ij}^{(k)}\in G_k$ for $k=0,1,\ldots,n-1$.

\begin{theorem}\label{LYthm2.1} {\rm \cite[Theorem 2.1]{LY}
Suppose $\Bbb F_{2^n}=\Bbb F_{2}(w)$, where $w$ is a root of an irreducible polynomial of degree $n$ over $\Bbb F_2$. Let $C$ be an $[l,k]$ linear code over $\Bbb F_{2^n}$ with generator matrix $G:=G_{0}+wG_{1}+\cdots+w^{n-1}G_{n-1}$, where $G_i \in M_{k\times l}(\Bbb F_{2})$. 
Then the binary subfield code $C^{(2)}$ with respect to $C$ has a generator matrix  
\[ G^{(2)}=\left(\begin{array}{c} G_{0}\\ 
G_{1}\\
\vdots \\
G_{n-1} \end{array}\right).\]

Moreover, if the code $C_{D}$ has the defining set $D=D_{0}+wD_{1}+\cdots+w^{n-1}D_{n-1}$ with $D_{i} \subseteq \Bbb F_2^m$, $0\leq i \leq n-1$, then the binary subfield code $C_{D}^{(2)}$ with respect to $C_{D}$ has the defining set: 
\[ D^{(2)}=\{(\mathbf d_{0}, \mathbf d_{1},\cdots,\mathbf d_{n-1}): \mathbf d_{i}\in D_{i},\, 0\leq i \leq n-1 \}.\]
}
\end{theorem}


\begin{theorem}{\rm \cite[Theorem 5.1]{LY}\label{LYthm5.1}
Let $L_i$ be nonempty subsets of $[m]$ and let $D = \Delta_{L_0} + w\Delta_{L_1} + \cdots + w^{n-1}\Delta_{L_{n-1}} \subseteq \Bbb F_{2^n}^m$. Then the subfield code $C_{D^*}^{(2)}$ is a $1$-weight linear code over $\Bbb F_{2^n}$ of length $2^{\sum_{i=0}^{n-1}|L_i|}-1$, dimension $\sum_{i=0}^{n-1}|L_i|$, and distance $2^{\sum_{i=0}^{n-1}|L_i|-1}$. It is a Griesmer code and hence distance optimal. Further, it is a minimal code.
}
\end{theorem}

\begin{theorem}{\rm\cite[Theorem 5.2]{LY}\label{LYthm5.2}
Let $L_i$ be nonempty subsets of $[m]$ and let $D = \Delta_{L_0} + w\Delta_{L_1} + \cdots + w^{n-1}\Delta_{L_{n-1}} \subseteq \Bbb F_{2^n}^m$. Then the subfield code $C_{D^c}^{(2)}$ is a $2$-weight binary linear code over $\Bbb F_{2}$ of length $2^{nm}-2^{\sum_{i=0}^{n-1}|L_i|}$, dimension $nm$, and distance $2^{nm-1}-2^{\sum_{i=0}^{n-1}|L_i|-1}$. In particular, $C_{D^c}^{(2)}$ has nonzero codewords of weights $2^{nm-1}-2^{\sum_{i=0}^{n-1}|L_i|-1}$ and $2^{nm-1}$. It is a Griesmer code and hence distance optimal. Further, it is a minimal code if $2^{\sum_{i=0}^{n-1}|L_i|} \leq nm-2$.
}
\end{theorem}

Let $\mathcal{R}_{2}:=\mathbb{F}_{2}[x]/\langle x^{3}-x\rangle $. Then we have the following results.

\begin{theorem}{\rm \cite[Theorem 3.2]{BSS}
Let $\mathcal{B}=\{\boldsymbol{e}_1=1+u^2<\boldsymbol{e}_2=u^2<\boldsymbol{e}_3=u+u^2\}$ be an ordered basis for $\mathcal{R}_{2}$ over $\mathbb{F}_{2}$. Suppose $G:=\boldsymbol{e}_{1}G_{1}+\boldsymbol{e}_{2}G_{2}+\boldsymbol{e}_{3}G_{3}\in M_{k\times n}(\mathcal{R}_{2})$ generates the linear code $\mathcal{C}$ over $\mathcal{R}_{2}$, where $G_{i}\in M_{k\times n}(\mathbb{F}_{2}),1\leq i\leq3$. Then the subfield code $\mathcal{C}^{(2)}$ of $\mathcal{C}$ is linear over $\mathbb{F}_2$ and is generated by
$$G^{(2)}=\left(\begin{array}{cc}G_1\\G_2+G_3\\G_2\end{array}\right).$$

Moreover, $\mathcal{C}_{D}^{(2)}=\mathcal{C}_{D^{(2)}}$ , that is, if the defining set of the code $\mathcal{C}_{D}$ is $D=\boldsymbol{e}_1D_1+\boldsymbol{e}_2D_2+\boldsymbol{e}_3 D_3\subseteq \mathcal{R}_{2}^m$ , where for $D_{i}\subseteq\mathbb{F}_{2}^{m},1\leq i\leq3$, the defining set of $\mathcal{C}_{D}^{(2)}$ is
$$D^{(2)}=\{(d_1,d_2+d_3,d_2):d_i\in D_i,1\leq i\leq3\}.$$
}
\end{theorem}

\begin{theorem}{\rm\cite[Theorem 4.1]{BSS}\label{BSSthm4.1}
Let $L,M,N\subseteq \{1,2,\ldots,m\}$, $\mathcal{R}_{2}$ be $\Bbb F_2$-valued trace of the $\Bbb F_2$-algebra $\mathcal{R}_{2}:=\mathbb{F}_{2}[x]/\langle x^{3}-x\rangle $, and the defining set $D:=(1+u^2)D_1+u^2D_2+(u+u^2)D_3 \subseteq \mathcal{R}_{2}^m$, where $u=x+\langle x^{3}-x\rangle$, $D_{1}\in\{\Delta_{L},\Delta_{L}^{c}\}$, $D_{2}\in\{\Delta_{M},\Delta_{M}^{c}\}$ and $D_{3}\in\{\Delta_{N},\Delta_{N}^{c}\}$. 
In various scenarios of selecting different defining sets, the authors construct several optimal binary subfield codes $\mathcal{C}_{D}^{(2)}$, with some being optimal under certain conditions. 

$(1)$ $D=\boldsymbol{e}_1\Delta_L+\boldsymbol{e}_2\Delta_M+\boldsymbol{e}_3\Delta_N\subseteq\mathcal{R}_2^m$. Then $\mathcal{C}_{D}^{(2)}$ is a $1$-weight binary linear code with parameters $[2^{|L|+|M|+|N|},|L|+|M|+|N|,2^{|L|+|M|+|N|-1}]$. The $\mathcal{C}_{D}^{(2)}$ meets the Griesmer bound and hence it is distance-optimal.

$(2)$ $D=\boldsymbol{e}_{1}\Delta_{L}^{c}+\boldsymbol{e}_{2}\Delta_{M}+\boldsymbol{e}_{3}\Delta_{N}\subseteq\mathcal{R}_{2}^{m}$. Then $\mathcal{C}_{D}^{(2)}$ is a $2$-weight binary linear code with parameters $[(2^m-2^{|L|})\times2^{|M|+|N|},m+|M|+|N|,(2^m-2^{|L|})\times2^{|M|+|N|-1}]$. The $\mathcal{C}_{D}^{(2)}$ meets the Griesmer bound and hence it is distance-optimal.

$(3)$ $D=\boldsymbol{e}_1\Delta_L+\boldsymbol{e}_2\Delta_M^c+\boldsymbol{e}_3\Delta_N\subseteq\mathcal{R}_2^m$. Then $\mathcal{C}_{D}^{(2)}$ is a $2$-weight binary linear code with parameters $[(2^m-2^{|N|})\times2^{|L|+|M|},m+|L|+|M|,(2^m-2^{|N|})\times2^{|L|+|M|-1}]$. The $\mathcal{C}_{D}^{(2)}$ meets the Griesmer bound and hence it is distance-optimal.

$(4)$ $D=\boldsymbol{e}_1\Delta_L+\boldsymbol{e}_2\Delta_M+\boldsymbol{e}_3\Delta_N^c\subseteq\mathcal{R}_2^m$. Then $\mathcal{C}_{D}^{(2)}$ is a $2$-weight binary linear code with parameters $[(2^{m}-2^{|N|})\times2^{|L|+|M|},m+|L|+|M|,(2^{m}-2^{|N|})\times2^{|L|+|M|-1}]$. The $\mathcal{C}_{D}^{(2)}$ meets the Griesmer bound and hence it is distance-optimal.

$(5)$ $D=\boldsymbol{e}_1\Delta_L^c+\boldsymbol{e}_2\Delta_M^c+\boldsymbol{e}_3\Delta_N\subseteq\mathcal{R}_2^m$. Then $\mathcal{C}_{D}^{(2)}$ is a $4$-weight binary linear code with parameters $[(2^m-2^{|L|})\times(2^m-2^{|M|})\times2^{|N|},2m+|N|,(2^m-2^{|L|}-2^{|M|})\times2^{m+|N|-1}]$. The $\mathcal{C}_{D}^{(2)}$  is distance-optimal if $2^{|L|+|M|+|N|}\leq2(m-1)+|N|$. 

$(6)$ $D=\boldsymbol e_{1}\Delta_{L}^{c}+\boldsymbol e_{2}\Delta_{M}+\boldsymbol e_{3}\Delta_{N}^{c}\subseteq\mathcal{R}_{2}^{m}$. Then $\mathcal{C}_{D}^{(2)}$ is a $4$-weight binary linear code with parameters $[(2^m-2^{|L|})\times(2^m-2^{|N|})\times2^{|M|},2m+|M|,(2^m-2^{|L|}-2^{|N|})\times2^{m+|M|-1}]$. The $\mathcal{C}_{D}^{(2)}$  is distance-optimal if $2^{|L|+|M|+|N|}\leq2(m-1)+|M|$.

$(7)$ $D=\boldsymbol{e}_{1}\Delta_{L}+\boldsymbol{e}_{2}\Delta_{M}^{c}+\boldsymbol{e}_{3}\Delta_{N}^{c}\subseteq\mathcal{R}_{2}^{m}$. Then $\mathcal{C}_{D}^{(2)}$ is a $4$-weight binary linear code with parameters $[(2^m-2^{|M|})\times(2^m-2^{|N|})\times2^{|L|},2m+|L|,(2^m-2^{|M|}-2^{|N|})\times2^{m+|L|-1}]$. The $\mathcal{C}_{D}^{(2)}$  is distance-optimal if $2^{|L|+|M|+|N|}\leq2(m-1)+|L|$.

$(8)$ $D=\boldsymbol{e}_{1}\Delta_{L}+\boldsymbol{e}_{2}\Delta_{M}+\boldsymbol{e}_{3}\Delta_{N}\subseteq\mathcal{R}_{2}^{m}$. Then $\mathcal{C}_{D^{c}}^{(2)}$ is a $2$-weight binary linear code with parameters $[(2^{3m}-2^{|L|+|M|+|N|}),3m,2^{3m-1}-2^{|L|+|M|+|N|-1}]$. The $\mathcal{C}_{D^{c}}^{(2)}$ meets the Griesmer bound and hence it is distance-optimal.
}
\end{theorem}

For subfield codes, the following problem is interesting and is valuable to consider. 

\begin{problem} {\rm 
Find more  new codes in the sense that they are inequivalent to the currently best-known linear codes presented in \cite[Table 11]{BSS}.

}
\end{problem}

\section{Concluding Remarks}

In this paper, we have explored various codes over finite rings and finite fields constructed from simplicial complexes and even posets, along with their related Gray images and subfield codes. We have also highlighted some open problems in this area, inviting readers to engage with these challenges. It would be nice if these problems are proved or disproved.

It is worth noting that the concept of simplicial complexes has been further generalized in a recent work \cite{H23}. This may be another interesting research direction  in coding theory.

\bigskip
\section*{Acknowledgments}


\end{document}